# Revisiting the Schottky-Mott Relation for Metal/ Transition Metal Dichalcogenide Interfaces: The Role of Interface Dipoles

Chih-Piao Chuu[1*], Shu-Jui Chang[1], Yu-Miin Sheu[3], Zih-Syuan Huang[1,2], Wei-Tung Liu [3], Po-Wei Chen[3], Sui-An Chou[1], Yu-Tung Lin[1,2], Chih-I Wu[2], Wen-Hao Chang[3], Chun-Liang Lin[3], H.-S. Philip Wong[4], and Iuliana P. Radu[1*]

[1] Corporate Research, Taiwan Semiconductor Manufacturing Company, Hsinchu 30091, Taiwan,

[2] Graduate Institute of Photonics and Optoelectronics, National Taiwan University, Taipei 106, Taiwan,

[3] Department of Electrophysics, National Yang Ming Chiao Tung University, 1001 Ta Hsueh Road, Hsinchu 30050, Taiwan;

[4] Department of Electrical Engineering and Stanford SystemX Alliance Paul G. Allen Building 312X 420 Via Palou, Stanford University, Stanford, California 94305, United State;

*Corresponding author: *Email: cpchuu@tsmc.com (C.-P. Chuu); Email: IRADU@tsmc.com (I. P. Radu)*

## Abstract

High contact resistance, predominantly stemming from Fermi-level pinning (FLP), becomes a critical bottleneck in the contact engineering of transition metal dichalcogenides (TMDs) based two-dimensional field-effect transistors (2DFETs). Our study re-examines the inherent nature of FLP in 2D semiconductors, particularly focusing on regimes where the material thickness is commensurate with the interface thickness. Through the synergistic application of first-principles calculations and experimental validation via ultraviolet photoelectron spectroscopy (UPS) and scanning tunneling spectroscopy (STS), we rigorously establish the metal effective work function (eWF) as the definitive figure of merit for substantiating the Schottky-Mott relation. Our findings indicate that FLP is intrinsically governed by both long-range dipole interactions and extended short-range Lennard-Jones potentials. This framework, in conjunction with the linear Schottky-Mott relation, improves predictive capabilities in contact engineering.

Herbert Kroemer's renowned statement, "the interface is the device,"[1] became a reality when Geim and Novoselov's group first mechanically exfoliated single-layer graphene from graphite.[2] This landmark achievement paved the way for the fundamental studies and device fabrications of other 2D materials, such as $MoS_2$, $WS_2$, and $WSe_2$, in monolayer form, illuminating new possibilities for extending Moore's law. Compared to silicon, their atomic thin physical thickness enhances the electrostatic control over the channel and allows for further gate length scaling. However, this ultra-slim body is a double-edged sword: while offering advantages, it also renders these materials inevitably susceptible to environmental influences. One of the critical bottlenecks hindering the further advancement of 2D FETs is their high contact resistance—particularly in p-type FETs, often attributed to Fermi level pinning (FLP). FLP is quantified by the pinning factor ($S=|d\Phi_{SBH}/dW_M|$), which measures the change in carrier barrier height $\Phi_{SBH}$ relative to the metal work function $W_M$. The intrinsic FLP at a metal-semiconductor junction can originate from three primary mechanisms: (1) Surface states from dangling bonds pinning the Fermi level in the bandgap (Bardeen's theory)[3]; (2) Metal-Induced Gap States (MIGS) due to evanescent metal wave functions penetrating the semiconductor (Heine's model)[4]; or (3) interface dipoles that create an 'effective' metal work function through dipole screening.[5] For van der Waals (vdW) materials, which inherently lack dangling bonds, the surface states contribution to FLP is negligible. Consequently, the intrinsic FLP in these systems is attributed either to MIGS, to interface dipoles, or to a combination of both mechanisms.[6-10]

Building upon this, previous studies on monolayer transition metal dichalcogenide (TMD) field-effect transistors have reported that S is typically around 0.1.[11-16] This phenomenon is largely attributed to chalcogen vacancies ($V_{S/Se}$). These defects could form during either the growth of 2D materials or metal contact deposition processes at high temperature. The formation energy (~1 eV) of chalcogen vacancies are relatively low compared to transition metal vacancies ($V_m$ ~3–8 eV)[17-19] or silicon vacancies ($V_{Si}$ ~4 eV). These defects introduce localized states characterized by sharp peaks in the density of states (DOS) below the conduction band minimum (CBM).[20]

While extrinsic factors, such as a substantial defect-induced density of states (DOS), can cause the Fermi level (FL) to become pinned at the defect level—a phenomenon robustly verified experimentally[16,21] —our focus here is on intrinsic Fermi Level Pinning (FLP). Intrinsic FLP, primarily governed by the interface between the top-layer TMD and metals,[22] constrains S to values below 0.4.[16] Notably, n-type Schottky contacts for $MoS_2$ devices are consistently observed across a broad spectrum of metals, encompassing both low and high work function (WF) materials.[23] In this paper, we shift our focus to the intrinsic nature of FLP, asserting that it is fundamentally governed by the interface dipole, which originates from long-range Coulomb interactions synergistically coupled with van der Waals (vdW) short-range forces. To move beyond the conventional metal WF, we introduce and adopt the effective work function (eWF), precisely defined as $W_{M/2D} = W_M - \Delta V_S$. This definition is rooted in the Schottky-Mott relation, with Schottky barrier heights (SBHs) derived from density functional theory (DFT) calculations. The eWF emerges as a compelling predictive figure of merit for understanding carrier injection barriers at metal/2D interfaces. Crucially, this eWF is not only measurable but also demonstrates a strong correlation with electron SBHs. Our theoretical framework finds compelling corroboration through experimental data obtained from Ultraviolet Photoelectron Spectroscopy (UPS) and Scanning Tunneling Spectroscopy (STS). Specifically, for monolayer $MoS_2$ and $WSe_2$ contact structures, these measurements largely affirm the predictions of the Schottky-Mott relation. This holistic approach empowers us to re-examine the Schottky-Mott rule with a quantifiable figure of merit, consequently enabling sophisticated predictive contact engineering and facilitating the judicious selection of optimal metals for achieving desired ohmic contact characteristics.

## ■ RESULTS AND DISCUSSION

### Interface Metrics: Work Function vs. Effective Work Function

Based on density functional theory (DFT) calculations, the calculated Schottky barrier heights (SBHs) at metal/TMD interfaces exhibit a dispersed relationship with the metal work function (WF) but a linear dependence on the effective work function (eWF), as depicted in Figure 1. This observed linearity with eWF is consistent with the Schottky-Mott relation, and independent of the choice of exchange functionals, yielding a pinning factor (S) of 1 (see Figure S1 and Table S1). Conversely, when employing the traditional metal WF, the SBH data points appear scattered, displaying low slopes of 0.40, 0.30, and 0.25 for monolayer $MoS_2$, $WS_2$, and $WSe_2$, respectively. This disparity underscores that eWF is the more appropriate figure of merit for investigating interface properties, as it inherently incorporates dipole effects.

### Electronic Structure and Band Alignment at Metal/TMD Interfaces

The work functions (WF) and effective work functions (eWF), as defined in Figures 2a–d, were calculated from the plane-averaged electrostatic Hartree potential.

Electron Schottky barrier heights (SBHs) were derived from the orbital-projected band structure of the metal-TMD stacked superstructure. Specifically, $\Phi_{SBH} = E_C - E_F$ at the K-valley, where the superstructure was constructed with a strain (applied to the metal) of less than 3%. At the K-valley, the conduction band minimum (CBM) is predominantly composed of $dz^2$ orbitals, whereas the valence band maximum (VBM) is primarily formed by $d_{x2-y2}/d_{xy}$ orbitals. Unlike the CBM, accurately determining the VBM's position presents a significant challenge. For metals such as Rh and Ru, substantial metal-induced gap states (MIGS) appear below the Fermi level. In the case of strongly binding metals like Ti and Hf, the valence band edge is either entirely submerged within a dense MIGS 'sea' (e.g., Ti) or exhibits significantly distorted band dispersion (e.g., Hf), see Figure S2. Consequently, the hole SBHs presented in Figure 1 are obtained by subtracting the electron SBHs from the fundamental band gap of monolayer TMDs, which is approximately 2.0 eV. Spin-orbit coupling (SOC) was not included in these calculations due to its negligible effect on the CBM at the K-valley. Our emphasis lies on the pinning factor (S, or slope) rather than the absolute SBH magnitude, as density functional theory (DFT) is known to underestimate the fundamental bandgap. Thus, the SBH values are accurate up to a constant offset. Strain effects, which are dependent on the contact metals, will be discussed in later sections, where the Schottky-Mott relation remains valid even with strain applied to the metal.

Extrapolated eWFs required for ohmic p-type contacts of monolayer $MoS_2$, $WS_2$, and $WSe_2$ are 6.2, 5.9, and 5.4 eV, respectively. These values are significantly higher (~1 eV) than the eWFs of the highest WF metals (e.g., Pt and Pd) considered in our simulations. For instance, the eWF of Pt on monolayer $MoS_2$, $WS_2$, and $WSe_2$ is 5.0 eV, 4.8 eV, and 4.5 eV, respectively. This clearly illustrates that p-FET on-state current is constrained by contact resistance, primarily due to the large Schottky barrier. Consequently, interface dipole engineering is crucial for achieving ohmic p-type contacts. Alternative contact strategies have been explored as prospective approaches for p-type contact engineering, including the selection of semimetal or 2D contacts to mitigate MIGS[24,25] and the theoretically predicted ultrahigh WF exceeding 7 eV on TiN(111).[26]

The band diagrams of the metal/TMD interface before and after contact are shown in Figure 2a for n-type contacts ($W_M < W_{2D}$) and Figure 2b for p-type contacts ($W_M > W_{2D}$). Due to the interface dipole, carriers in 2D materials side see a modified metal work function $W_{M/2D}$, herein called eWF, resulting in $W_M = W_{M/2D} + \Delta V_S$ at the interface with a distance of $d_{int}$. The quantities $W_M$, $W_{M/2D}$, and work function difference $\Delta V_S$ can be numerically calculated through electrostatic local potential across the normal surface from metal to 2D materials, for examples as shown in Figures 2c and 2d, for n-type contact of Sb and p-type contact of Pd. The Schottky barrier height $\Phi_{SBH}$, defined in the band diagram, satisfies $\Phi_{SBH} + \chi_{2D} = W_{M/2D}$. The electron affinity ($\chi_{2D}$) of freestanding monolayer $MoS_2$, $WS_2$, and $WSe_2$ can be calculated as 4.2 eV, 3.9 eV, and 3.5 eV, respectively. In metal/TMD heterostructures, the relationship $\chi_{2D} = W_{M/2D} - \Phi_{SBH}$ applies. As shown in Table 1, the derived $\chi_{2D}$ values are 4.2 eV, 3.9 eV, and 3.5 eV for monolayer $MoS_2$, $WS_2$, and $WSe_2$, respectively. Notably, these values remain consistent regardless of the specific metal contacts, agreeing excellently with those of freestanding monolayer cases. Furthermore, the electron affinity ($\chi_{2D}$) remains nearly unchanged even as the interlayer distance varies. This consistency indicates that the band diagrams presented in Figures 2a and 2b provide a good representation of the metal-TMD interface. At this interface, dipole charges reside (as depicted by the density difference in Figures 2c and 2d).

## Physics of the Interface Dipole

In general, interface dipoles increase the eWF of low WF metals and decrease the eWF of high WF metals, a phenomenon intrinsic to FLP. A simple parallel plate capacitor model provides a physics-based approach to understanding this dipole effect. Using Gauss's law, the dipole field, $E_{dip} = \sigma_S / \varepsilon_S$, and voltage drop ($\Delta V_S$) given by the work done by the dipole field across the interface is $\Delta V_S = e\sigma_S d_{int} / \varepsilon_S$, where $\sigma_S$ and $\varepsilon_S$ is the accumulated interface charges density and effective permittivity at interface, and the optimized interlayer distance $d_{int}$ is determined at the maximum binding energy in Figures 3a and 3b. Even though both $\sigma_S$ and $\varepsilon_S$ are experimentally challenging to determine, estimates ($\varepsilon_S \sim \varepsilon_0$) indicate a significant $\Delta V_S$ of ~0.5 eV for practical interface charge densities of $10^{13}$ cm$^{-2}$ across atomic-scale interlayer distances (~2.5 Å). The interface dipole, $\mathbf{P} = Q_S \mathbf{d}$, diminishes away at large $|\mathbf{d}|$, suggesting effective charges ($Q_S = \sigma_S A$) at interface also reduces to zero. It is reasonable to assume that $Q_S$ originates from charge transfer associated with interface bonding, which vanishes when the atomic wavefunction overlap becomes negligible. At equilibrium, $\mathbf{P}/A = \sigma_S d_{int} = (\Delta V_S / e)\varepsilon_S$ is proportional to work function difference $\Delta W = W_M - W_{M/2D}$.

## Interlayer Interaction and Binding

The equilibrium interlayer distance ($d_{int}$) in metal/TMD heterostructures is determined by the balance between attractive Lifshitz-van der Waals (LW) interactions and repulsive Born interactions. For simplicity, we employ the Hamaker approach[27] to model the attractive interaction. Specifically, for any particle on the TMD surface at a distance r from the metal surface, an effective pairwise interaction can be described by a power law: $V(r) = - C/r^6$, where C is defined as the Lifshitz-van der Waals (LW) potential coefficient. The total van der Waals (vdW) interaction between the metal and TMD can then be calculated by integrating this pairwise interparticle potential, multiplied by the total number of particles, over the entire volume of the metal.

$$V_{LW} = \int_d^{\infty} \int_0^R \frac{-C}{(x^2 + R^2)^3} \rho(2\pi R) dR dx = -\frac{C\rho\pi}{6} \frac{1}{d^3}$$

with ρ is the particle density of the metal, and d is the interlayer distance.

In Figure 3a, the binding energy $E_b$ between a monolayer $MoS_2$ and various overlayers were calculated. In most cases of overlayers (see Table 2 and the discussion accompanying Figure S3), the DFT calculated binding energy can be fitted using the modified Lennard-Jones (LJ) potential:

$$V_{int}^{LJ}(d) = 4E_{b,min}\,[(d_0/d)^{2n} - (d_0/d)^{n}]$$

where $E_{b,min}$ is the minimum binding energy evaluated at optimized interlayer distance $d_{int}=2^{1/n}d_0$, and $d_0$ is the distance where potential intersecting at $V_{int}^{LJ}=0$. As shown in Table 2, the attractive interaction's power law varies approximately as $1/d^3$, i.e. $n\simeq3$, which is consistent with the Hamaker approach, and insensitive to exchange functionals, see Figure S3 for more details. Figure 3b illustrates the minimum binding energy for different overlayers. Furthermore, the calculated binding energy, or interface adhesion energy (IAE), for bilayer $MoS_2$, *h*BN/$MoS_2$, and graphene/$MoS_2$ exhibits good agreement with experimental data.[28]

## Experimental Validation and Role of Defects

As shown in Figure 1, the Schottky-Mott relation, $\partial\Phi_{SBH}/\partial W_{M/2D}=1$, holds for a wide range of metals, with IAE ranging from 0.3 $J/m^2$ to 1.6 $J/m^2$, as calculated in Figure 3b. Our objective was to verify the validity of the Schottky-Mott relation for CVD-grown TMD samples. This is crucial because our simulations primarily considered pristine TMD. To achieve this, we prepared electron beam deposited metals on TMD/sapphire. These samples were then transferred from the sapphire to a Si substrate using Ag glue. This process yielded a TMD (0.7 nm)/metal (30 nm)/Ag/Si structure. This structure was then used for photoelectron spectroscopy (PES) measurements. The eWFs were measured using an X-ray source (XPS). Defects and bonding characteristics were studied using a UV source (UPS). During sample preparation, the deposited metal overlayer demonstrated higher susceptibility to strain compared to the underlying TMD layer. This observation is consistent with theoretical assumptions.

The scanning tunneling spectroscopy (STS) measurements were performed on same samples as above, to determine electron Schottky barriers (SBH), see Figure 4a. The extracted SBHs of metal contacts versus eWF and WF were plotted in Figure 4b and listed in Table 3. For $MoS_2$, the SBH has a pinning factor of S = 1.12 against the eWF, and S = 0.54 against the WF. For $WSe_2$, S = 1.22 against the eWF, and S = 0.8 against the WF. Based on four studied contact metals (Bi, Au, Pd, Pt), the eWF provides a better description of interface properties by accounting for the dipole-induced potential shift. For $WSe_2$, the correlation coefficient of the linear fit between SBH and eWF is relatively low at 0.44, compared to $MoS_2$, which has a value of 0.87. The strain effect was ruled out as a possible factor influencing the Schottky-Mott relation. This conclusion was reached after further DFT calculations on metal/TMD superstructures with varying strain confirmed that the linearity of the Schottky-Mott relation is insensitive to strain, as shown in Figure 4c and Table 4. Figure 4d compares the DFT-calculated ($\Phi_{DFT}$) and STS-measured ($\Phi_{STS}$) electron Schottky barrier heights. While good agreement is observed for $MoS_2$ systems (except Bi), deviations are evident for $WSe_2$ systems, particularly with Au, Pd, and Pt contacts. Furthermore, FLP is observed for high work function contact metals on $WSe_2$ samples, a phenomenon not evident in $MoS_2$ samples (refer to Figure S4). High-resolution XPS spectra of $WSe_2$ were acquired for both pristine samples and after metal deposition, as illustrated in Figure 5a. A shift in core levels is apparent after metal contact, which can be attributed to interlayer interaction. Notably, in Au/$WSe_2$ and Pd/$WSe_2$ samples, additional doublets are observed. These doublets serve as evidence of selenium (Se) defect formation, a phenomenon previously reported in other studies.[29-32] and was identified in the moire structures of Au/$WSe_2$ and Pd/$WSe_2$ samples, as shown in Figure S5.

The influence of Se vacancies was further investigated by calculating the band structures of Au/$WSe_2$ (R9/R7), Pd/$WSe_2$ (R13/R9), and Pt/$WSe_2$ (R13/R9) superstructures incorporating defective $WSe_2$ with Se defect concentrations of 7.1%, 5.5%, and 5.5%, respectively. For pristine $WSe_2$, the SBH for electrons at the K valley is found to be 0.77 eV, 1.01 eV, and 1.03 eV for Au, Pd, and Pt contacts, respectively. With the introduction of Se defects, the electron barrier for Au/$WSe_2$, Pd/$WSe_2$, and Pt/$WSe_2$ shifts to 0.89 eV, 0.90 eV, and 0.95 eV, respectively, indicating that the Fermi level becomes pinned. Additionally, a defect-induced band is observed at 0.25 eV, 0.30 eV, and 0.53 eV below the conduction band minimum, as depicted in Figure 5b. An increase in Se defect concentration leads to the calculated effective work function (eWF) for Pt/$WSe_2$, Pd/$WSe_2$, and Au/$WSe_2$ consistently converging to approximately 4.49 eV (Figure S6). This convergence implies that extrinsic Fermi Level Pinning (FLP) is primarily driven by Se defects. Beyond the S/Se defect which we identify as a main cause for deviations in $WSe_2$ data (Figure 4d), other factors may also contribute. We note that the calculated eWF and SBH may depend on functionals (see Figure1 and Figure S1 for D3 and optB86 exchange functionals, respectively), along with inherent DFT limitation like band gap underestimation can also influence the calculated SBHs. The residues of photo-resistance at interface could cause interface contamination or charge accumulation and therefore alter the interface dipole and eWF.

## Scaling Limits, Dimensionality, and Suppression of Fermi Level Pinning

In Figures 6 a–c show the dependence of eWF, WF, and SBH of contact metals on monolayer TMD systems for Sb/$MoS_2$ (R4/R7), Pd/$MoS_2$ (R4/R3), and Sb/$WSe_2$ (R4/R7) in an interlayer distance ranging from 1.6 to 6.0 Å. The effect of the interface dipole diminishes at distances greater than 5 Å, where WF ≈ eWF, indicating the Schottky limit. Below 5 Å, the SBH follows the trend of the eWF, suggesting that the eWF predominantly governs the electronic properties of the interface. When the interlayer distance is reduced below the optimal value ($d_{int} < d_{opt}$ - 0.5 Å), eWF and SBH become insensitive to further changes in distance, corresponding to FLP. This behavior can be interpreted as an increase in interface charge due to stronger hybridization of the metal *d*-orbitals at shorter interlayer distances, referred to the Bardeen limit. In contrast, WF exhibits quantum oscillations as the Fermi level crosses the peaks of density of states of quantum well states of quasi 2D electron gas.[26,33] The 2D electron affinity ($\chi_{2D}$) of monolayer $MoS_2$ and $WSe_2$, derived from the formula $\chi_{2D}$ = eWF-$\Phi_{SBH}$ in band diagram is presented in Figure 6d. This value shows little variation with contact metals or interlayer distance.

It is noteworthy that FLP persists even in metal-2D-metal sandwiched structures. For example, the electron SBH value slightly changes from 0.738 (Pd/$MoS_2$) to 0.742 eV (Pd/$MoS_2$/Pd), and from 0.096 (Sb/$MoS_2$) to 0.102 eV (Sb/$MoS_2$/Sb). This phenomenon holds true for both strong-binding contact metals (Pd), and weak-binding metals (Sb). In these capacitor-like structures, even though the net dipole field cancels out, the dipole-induced potential remains constant, thus maintaining the FLP. Furthermore, the presence of MIGS doubles, causing the system to approach closer to the Bardeen limit.

We also compared the calculated SBHs of Pd/$MoS_2$ under different substrates, including vacuum, Pd, hBN, $Al_2O_3$, and $HfO_2$, as presented in Figure 6e. The SBHs exhibited a change of less than 5% change across these various substrates. This suggests that the metal/TMD interface itself is the primary limiting factor determining the SBH.

While 2D interface dipole has intrinsic limitation in suppressing FLP, reducing the dimensionality of the interface dipole is shown to be an effective approach to suppress FLP. For instance, an edge contact structure made by completely etching the patterned area before metal deposition, exhibits nearly free FLP in Ni-1L $MoS_2$ edge contact transistor devices.[34] The interline (1D interface) of edge contact can be treated as an array of 1D line dipole with charge density of $\lambda$, illustrated in Figure 7a. A point charge $\delta q=\lambda\delta x$ located at (x,0) along the interline contributes a local dipole potential of $\delta V=k(\lambda\delta x)d\cos\theta/(z^2+x^2)$ at (0,z), where d (~$d_{int}$) is the distance between opposite charge arrays with a dipole moment of $\delta P=d\ \delta q$, and k is the Coulomb's constant. By integrating over the interline from $-\pi$ to $\pi$ to sum up the local dipole contribution, we obtain the 1D dipole potential, $V_{1D}(z) = 2kd\lambda/z$, which inversely decreases with distance from the interline. In comparison, the top contact structure, described by a parallel capacitor model, has no electric field outside, resulting in a constant dipole potential. Consequently, the effect of FLP does not diminish with distance (see the middle and right panels (high WF metal) of Figure 7a. This is consistent with the observations demonstrated in the local density of states calculations shown in Figures 7b and 7c, where the 0.8 eV electron Schottky barrier height of the Pd top contact increases to 1.2 eV for the Pd edge contact, corresponding to an eWF of 5.4 eV as shown in Figure 1a. This value (5.4 eV) is very close to the calculated WF of a 4-layer Pd film (5.3 eV) in the unpinned situation ($W_M = W_{M/2D}$ when $\Delta V_S = 0$), indicating that FLP is indeed suppressed through the reduced dipole, consistent with experimental observations.

## ■ CONCLUSION

Through first-principles calculations and experimental verification, we have demonstrated that the effective work function (eWF) is the optimal metric for evaluating Schottky barrier heights in metal-TMD interfaces, explicitly accounting for the effects of interface dipoles. The model extends effectively to bilayer TMDs, holding true irrespective of stacking order (Figure S7 and Table S2), underscoring the model's robustness despite expected changes in vdW interactions and electrostatic screening with increased thickness. Unlike traditional WF metrics, eWF provides predictive insights for carrier injection barriers, enabling improved contact engineering prior to device fabrication.

Our findings validate the Schottky-Mott relation for CVD-grown monolayer $MoS_2$ samples. Further refinement is needed to address exceptions, such as observed FLP in $WSe_2$ metal contacts (e.g., Au and Pd) due to the generation of Se vacancies.

This highlights the urgency of innovative engineering approaches, such as reduced-dimensionality contacts, to mitigate its adverse impact on device performance. The predictive power of eWF elevates the Schottky-Mott model as a practical tool for advancing 2D FET design, particularly in the quest for high-performing p-type ohmic contacts suitable for next-generation electronics.

**Figures and Tables:**

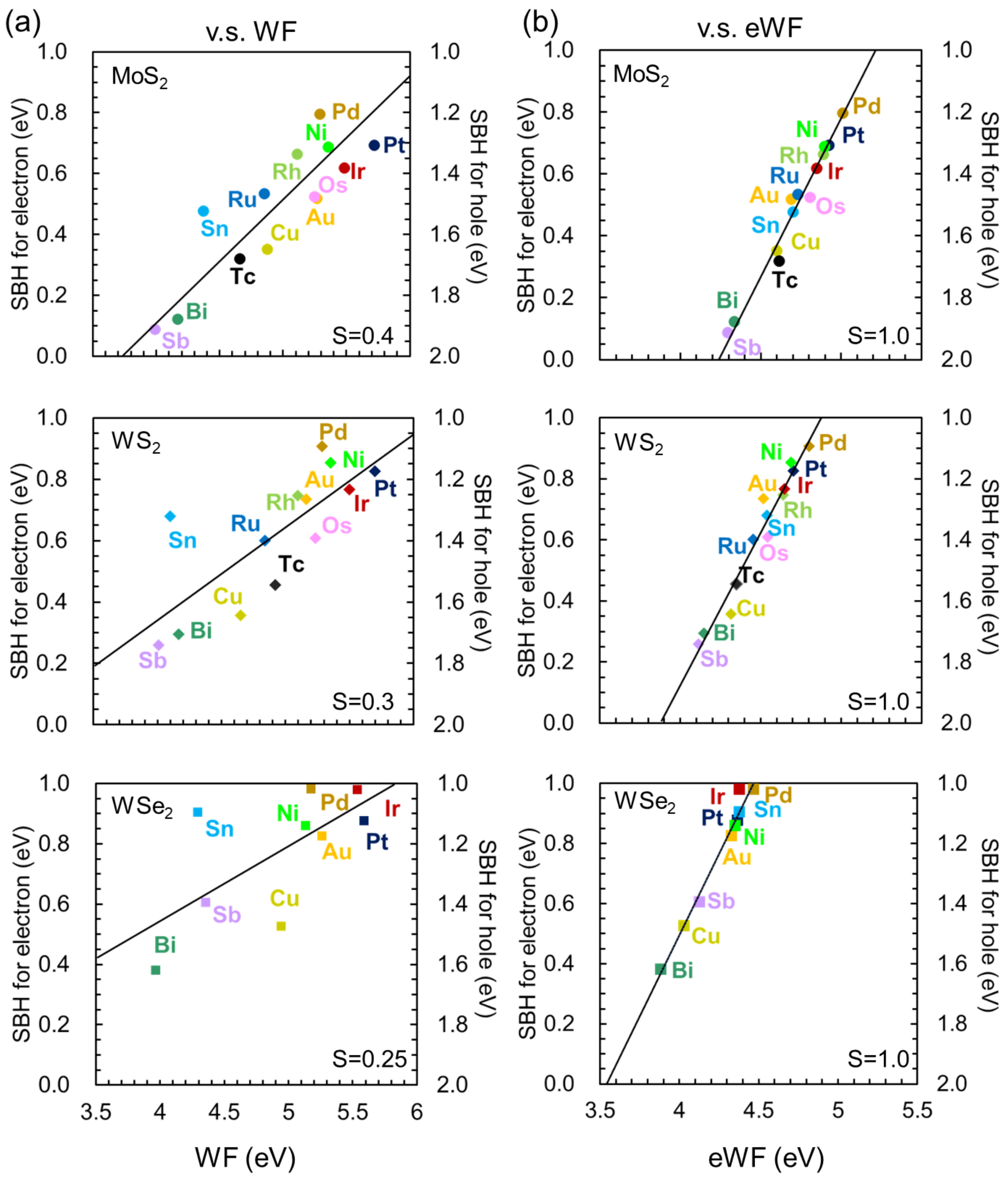

Figure 1. Calculated Schottky barrier heights (SBHs) of electrons v.s. (a) work function (WF), and v.s. (b) effective work function (eWF) of contact metals on monolayer $MoS_2$ (circles), $WS_2$ (diamonds), and $WSe_2$ (squares). Slope S was obtained from the linear fitting (solid line) of calculated data points of SBH versus WF/eWF for $MoS_2$, $WS_2$, and $WSe_2$.

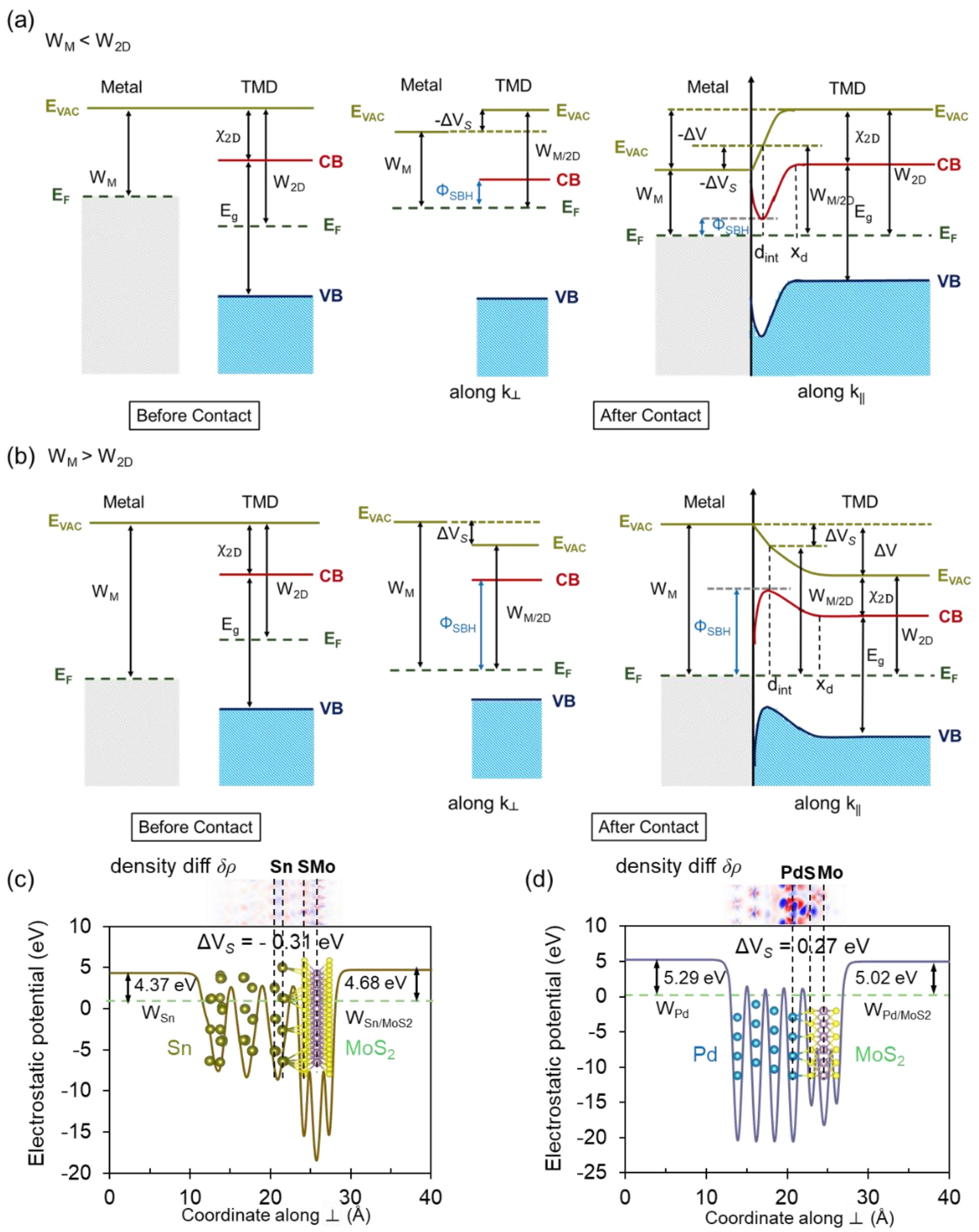

Figure 2. The band diagram of metal-TMD for (a) $W_M<W_{2D}$, and (b) $W_M>W_{2D}$ before (vacuum level $E_{VAC}$ is aligned) and after contact (Fermi level $E_F$ is aligned). The $W_M$ and $W_{M/2D}$ represent the work function and effective work function of metal, and $W_{2D}$, $\chi_{2D}$, $E_g$ is the work function, electron affinity, and bandgap of 2D semiconductor. $\Delta V_S$ and $\Delta V$ is the vacuum level difference at the interlayer distance $d_{int}$, and away the junction (> depletion length $x_d$) entering the flat band region. $\Phi_{SBH}$ is the electron Schottky barrier height. Normal to the surface, the carriers see quantized levels along $k_\perp$ in interlayer spacing, and band dispersion appears (>$x_d$) after carriers transport along $k_\parallel$ once source to drain bias is turned on. Electrostatic potential across the Schottky junction of (c) n-type Sn and (d) p-type Pd contact on monolayer $MoS_2$ reveal that the potential drop changes sign in two situations. The dipole charges reside at the interface were calculated by electron density difference before and after metal contact as $\delta\rho = \rho_{M/2D}-\rho_M-\rho_{2D}$.

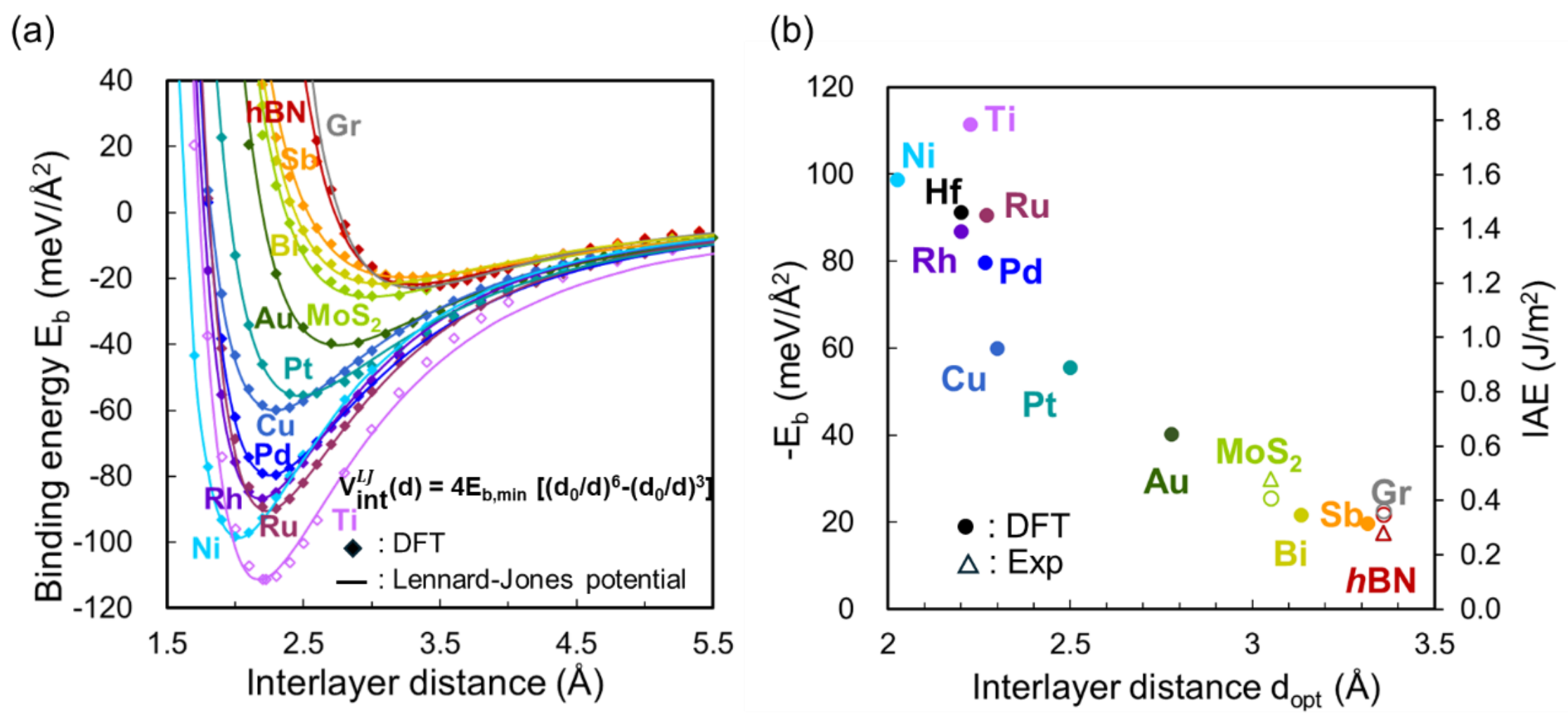

Figure 3. (a) Calculated binding energy $E_b$ (diamonds) of different overlayers and fitted 6-3 Lennard-Jones potential (solid lines) with respect to monolayer $MoS_2$ versus interlayer distance, and (b) calculated binding energy $-E_b$ of different overlayers on monolayer $MoS_2$ (circles) at optimized interlayer distance (energy minimum in 3(a)), compared with available experimental data (triangles).

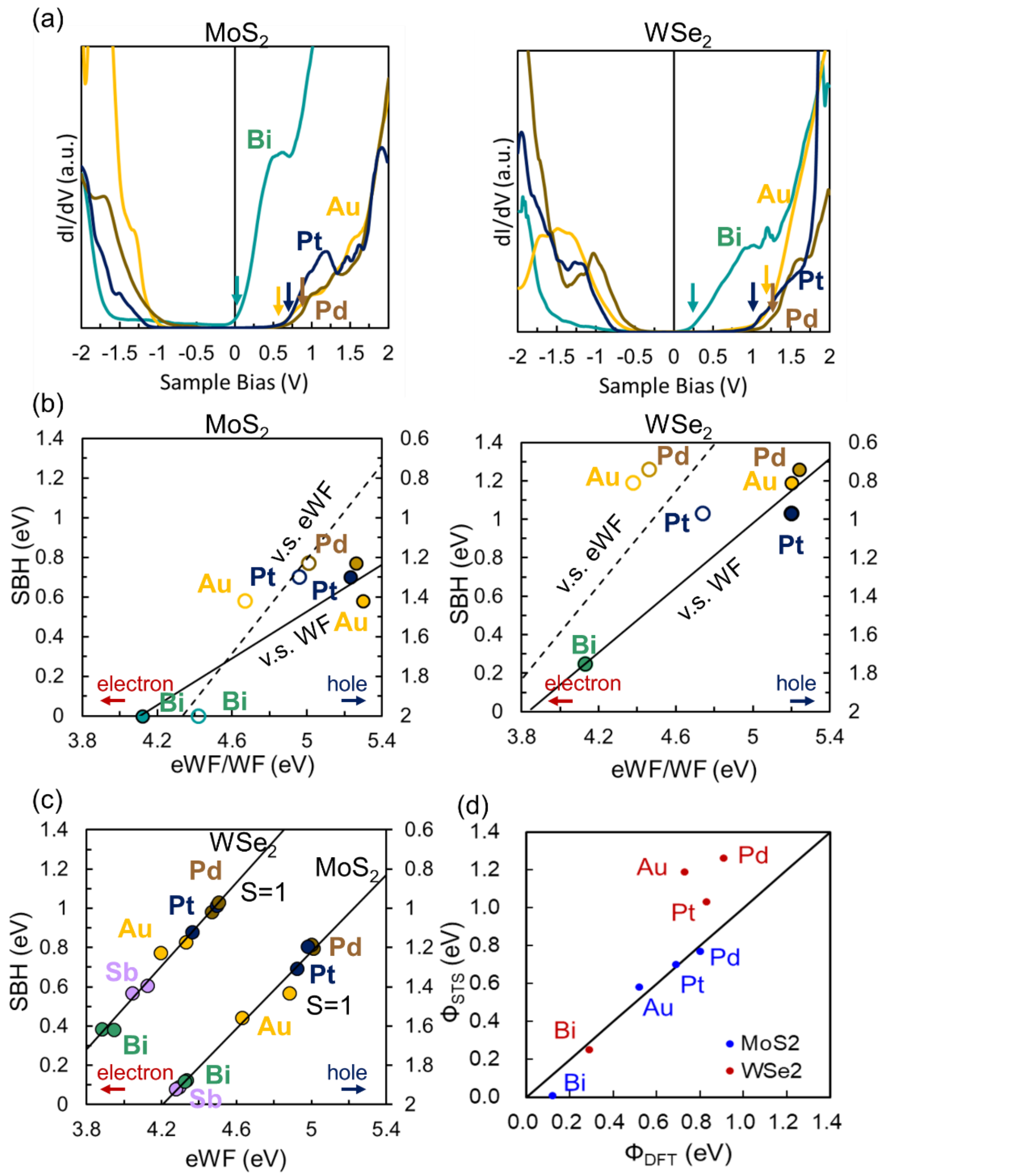

Figure 4. (a) STS measurements of monolayer $MoS_2$ (left) and $WSe_2$ (right) where arrows indicated the position of $E_C$ summarized in Table. 4, and (b) STS extracted SBHs versus UPS measured eWF (open) and WF (solid) of four contact metals of monolayer $MoS_2$ (left) and $WSe_2$ (right). (c) DFT calculated eWF and SBH of contact metals subject to different strains follow the same linear trend as predicted by the Schottky-Mott law, see Table. 3 for calculation details. (d) Comparison between DFT calculated electron Schottky barrier $\Phi_{DFT}$ and STS measured electron Schottky barrier $\Phi_{STS}$.

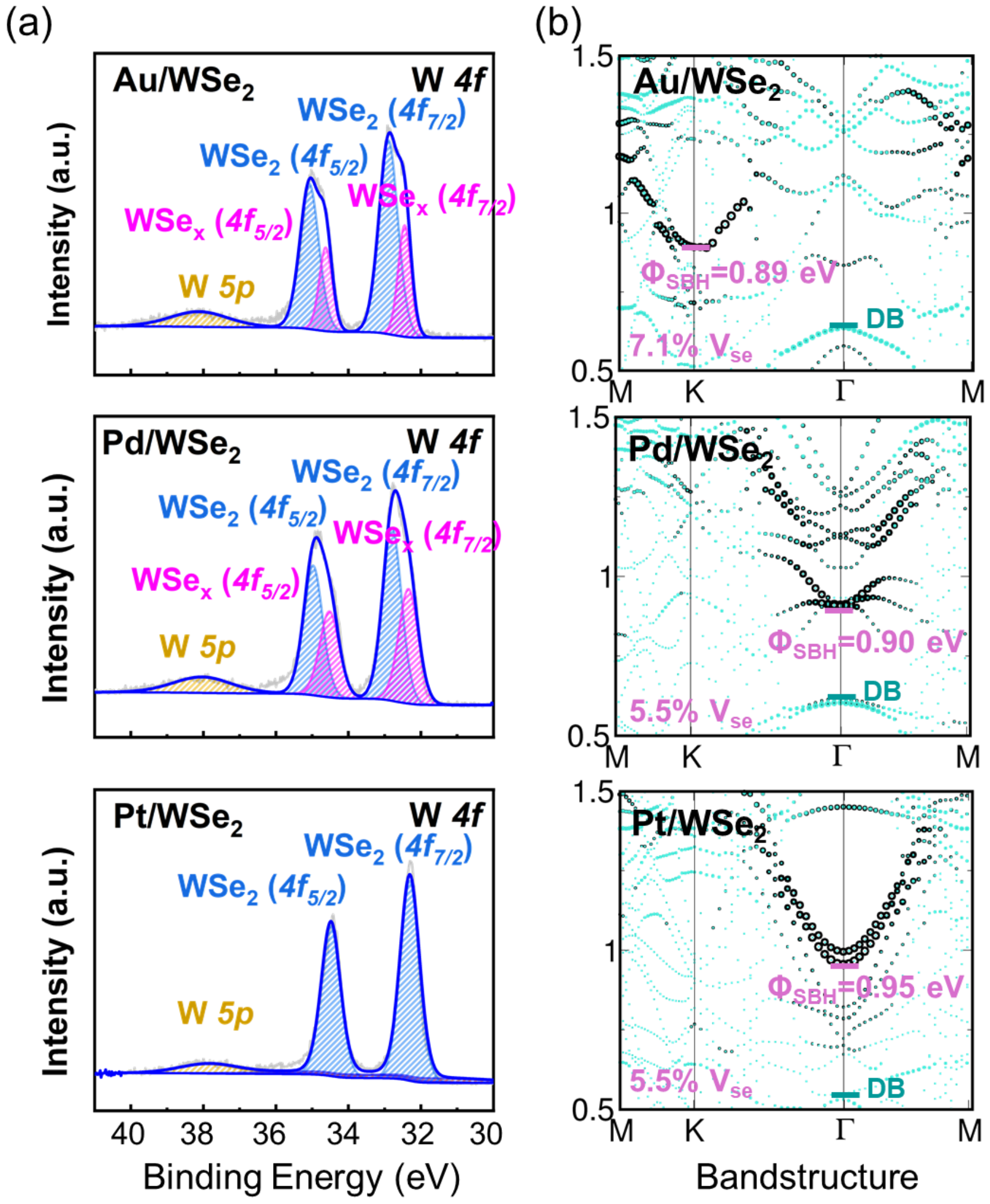

Figure 5. (a) XPS binding energy measurements of metal/$WSe_2$ samples. Additional peaks of $WSe_X$ were observed in Au/$WSe_2$ and Pd/$WSe_2$ samples due to the generation of Se defect. (b) DFT calculated mteal/defective $WSe_2$ bandstructures showing FLP is pinned among Au, Pd, and Pt contacts. DB indicates the defect induced band position.

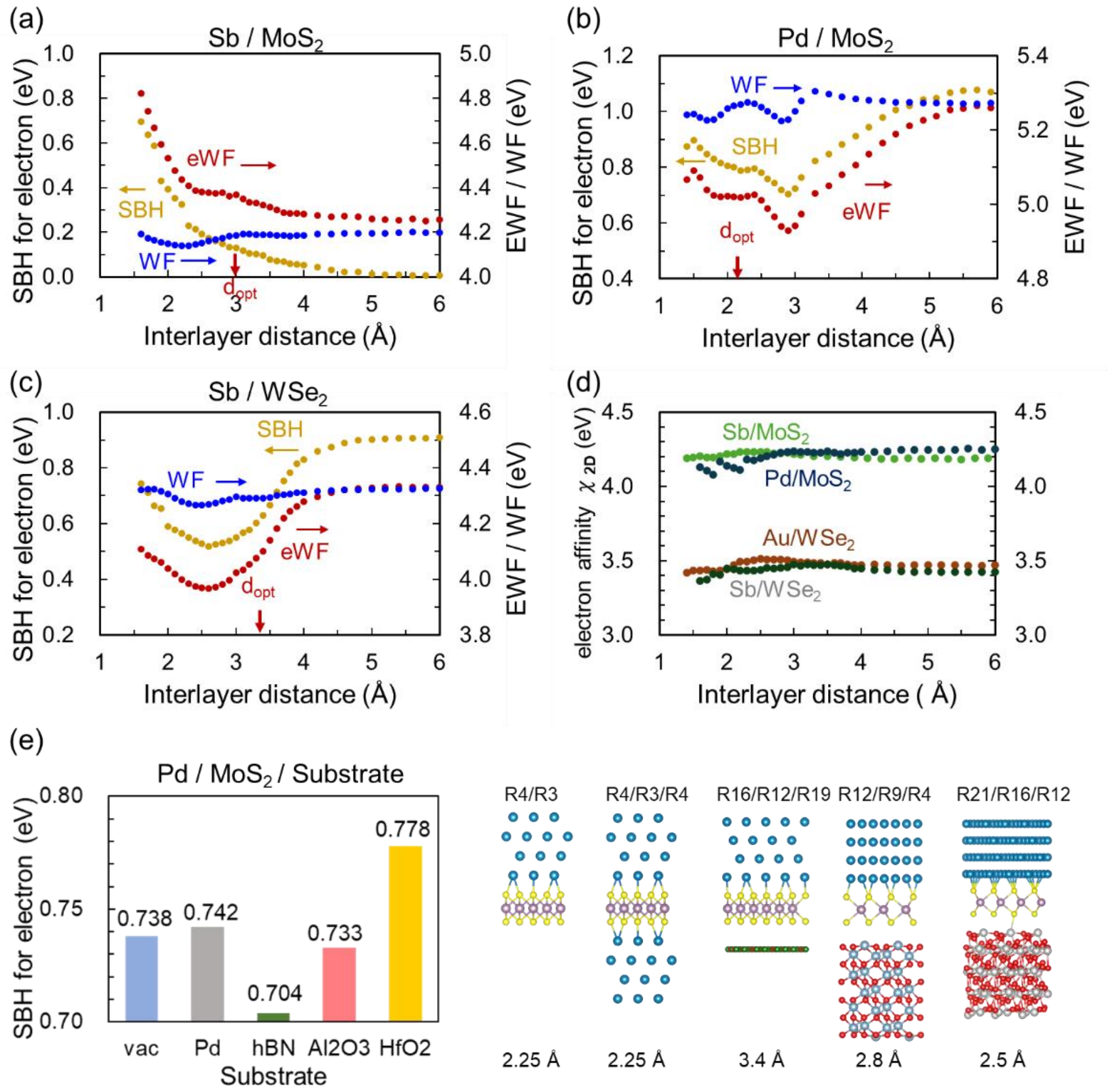

Figure 6. SBH for electrons (yellow), WF (blue), and eWF (red) as a functional of interlayer distance ($d_{Mo(W)S(Se)}$) dependent of (a) Sb/$MoS_2$ (R4/R7), (b) Pd/$MoS_2$ (R4/R3), (c) Sb/$WSe_2$ (R4/R7) by D3 exchange functionals, (d) Calculated electron affinity according to the formula, $\chi_{2D}$ = eWF- $\Phi_{SBH}$, for monolayer $MoS_2$ and $WSe_2$, (e) Calculated SBH for electron of Pd/$MoS_2$/vac(R4/R3), Pd/$MoS_2$/Pd(R4/R3/R4), Pd/$MoS_2$/hBN(R16/R12/R19), Pd/$MoS_2$/$Al_2O_3$(R12/R9/R4), and Pd/$MoS_2$/$HfO_2$(R21/R16/R12) of an interlayer distance from substrate listed in below.

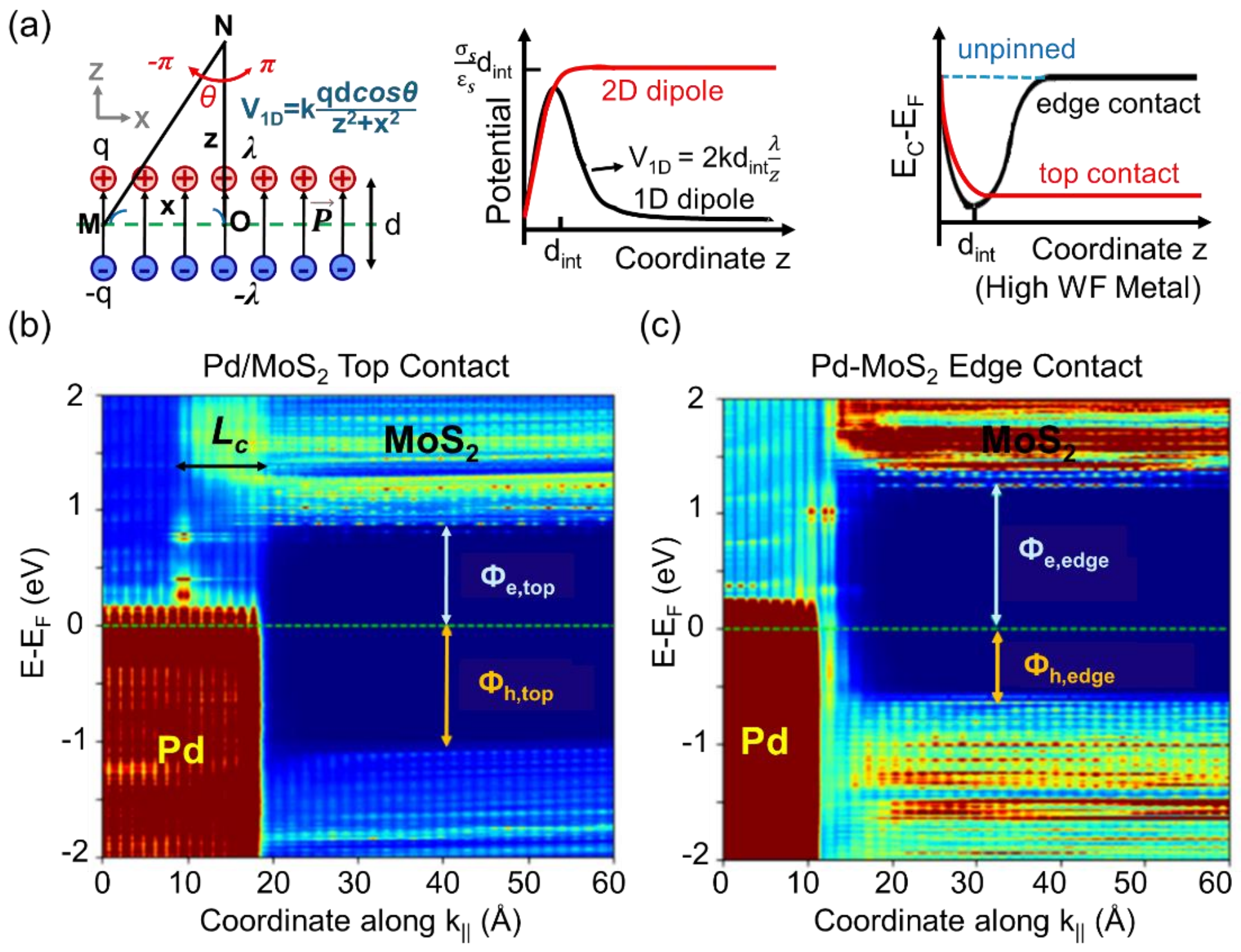

Figure 7. (a) (Left) Physics picture of 1D line dipole at interline (dashed line) of metal-2D edge contact structure with charge separation of distance $d \sim d_{int}$, the dipole potential at N is the sum over contribution from any local dipole moment at M with tilted angle $\theta$ through the whole interline. (Middle) Dipole potential of 1D dipole (black) and 2D dipole (red) varies along coordinate z away from the junction. (Right) Illustration of the effect of FLP for high WF metal in top (2D) and edge (1D) contact. (b) Local density of states (LDOS) of Pd top contact structure with 1 nm contact length $L_C$, and (c) Pd edge contact structure along transport direction $k_{||}$ on 1L $MoS_2$.

Table 1. Calculated interlayer distance $d_{int}$, effective work function (eWF) and electron Schottky barrier $\Phi_B$ of metal-monolayer TMD superlattices structure by D3 exchange functionals in Figure 1. The supercell (SC) structure $\sqrt{m} \times \sqrt{m}$ Metal /$\sqrt{n} \times \sqrt{n}$ TMD is represented by m/n for simplicity. The calculated lattice constant of monolayer $MoS_2$, $WS_2$, and $WSe_2$ is 3.165 Å, 3.175 Å, 3.30 Å, respectively.

| | $MoS_2$ | | | | $WS_2$ | | | | $WSe_2$ | | | |
|---|---|---|---|---|---|---|---|---|---|---|---|---|
| | SC | $d_{int}$ | eWF | $\Phi_B$ | SC | $d_{int}$ | eWF | $\Phi_B$ | SC | $d_{int}$ | eWF | $\Phi_B$ |
| Pt | 4/3 | 2.42 | 4.92 | 0.69 | 4/3 | 2.42 | 4.71 | 0.83 | 4/3 | 2.45 | 4.37 | 0.88 |
| Pd | 4/3 | 2.25 | 5.01 | 0.80 | 4/3 | 2.27 | 4.80 | 0.91 | 4/3 | 2.34 | 4.47 | 0.98 |
| Cu | 19/12 | 2.37 | 4.60 | 0.35 | 19/12 | 2.36 | 4.35 | 0.45 | 7/4 | 2.50 | 4.03 | 0.53 |
| Au | 9/7 | 2.84 | 4.69 | 0.52 | 9/7 | 2.83 | 4.52 | 0.73 | 4/3 | 2.85 | 4.33 | 0.83 |
| Ni | 7/4 | 2.00 | 4.90 | 0.69 | 7/4 | 2.09 | 4.69 | 0.85 | 7/4 | 2.23 | 4.35 | 0.86 |
| Bi | 9/19 | 3.10 | 4.34 | 0.12 | 9/19 | 3.14 | 4.15 | 0.29 | 12/13 | 3.27 | 3.88 | 0.38 |
| Sb | 4/7 | 3.23 | 4.30 | 0.09 | 4/7 | 3.36 | 4.11 | 0.26 | 4/7 | 3.35 | 4.13 | 0.61 |
| Sn | 9/19 | 2.90 | 4.68 | 0.48 | 9/19 | 2.90 | 4.54 | 0.68 | 12/13 | 3.10 | 4.38 | 0.91 |
| Ir | 4/3 | 2.29 | 4.85 | 0.62 | 4/3 | 2.30 | 4.65 | 0.77 | 13/9 | 2.47 | 4.38 | 0.98 |
| Rh | 4/3 | 2.19 | 4.89 | 0.66 | 4/3 | 2.18 | 4.64 | 0.75 | | | | |
| Ru | 4/3 | 2.24 | 4.73 | 0.53 | 4/3 | 2.20 | 4.46 | 0.60 | | | | |
| Os | 4/3 | 2.34 | 4.81 | 0.52 | 4/3 | 2.30 | 4.55 | 0.61 | | | | |
| Tc | 4/3 | 2.30 | 4.61 | 0.32 | 4/3 | 2.30 | 4.32 | 0.36 | | | | |

Table 2. Physical quantities of monolayer $MoS_2$ with different overlayers shown in Figure 3, superlattice ($MoS_2$/overlayer), lattice constant a, binding energy $E_b$ at interlayer distance $d_{int}$ calculated by optb86-vdW functionals, and power n of Lennard-Jones potential.

| overlayer | a (Å) | superlattice | $d_{int}$ (Å) | $E_b$ (meV/Å$^2$) | n |
|---|---|---|---|---|---|
| $MoS_2$ | 3.17 | R1/R1 | 3.05 | 25.5 | 3.0 |
| *h*BN | 2.52 | R12/R19 | 3.36 | 21.7 | 3.6 |
| Bi | 4.75 | R19/R9 | 3.13 | 21.6 | 3.0 |
| Sb | 4.54 | R13/R7 | 3.32 | 19.6 | 2.7 |
| Ni | 3.51 | R12/R19 | 2.03 | 98.6 | 3.2 |
| Cu | 3.64 | R9/R13 | 2.30 | 59.9 | 3.0 |
| Au | 4.16 | R16/R19 | 2.78 | 40.1 | 3.1 |
| Ru | 2.72 | R3/R4 | 2.27 | 90.5 | 3.3 |
| Rh | 3.83 | R3/R4 | 2.20 | 86.8 | 3.3 |
| Pd | 3.94 | R3/R4 | 2.27 | 79.6 | 3.1 |
| Pt | 3.97 | R3/R4 | 2.50 | 55.5 | 3.0 |
| Ti | 3.01 | R12/R13 | 2.23 | 111.3 | 3.1 |
| Hf | 3.20 | R1/R1 | 2.2 | 91.1 | 3.8 |

Table 3. UPS measured effective work function (eWF), work function (WF), and STS measured electron Schottky barrier ($\Phi_{STS}$) of monolayer $MoS_2$ and $WSe_2$ in Figures 4 (a)–(c).

| | $MoS_2$ | | | $WSe_2$ | | |
|---|---|---|---|---|---|---|
| | eWF | WF | $\Phi_{STS}$ | eWF | WF | $\Phi_{STS}$ |
| Bi | 4.42 | 4.12 | 0.006 | 4.13 | 4.13 | 0.25 |
| Au | 4.67 | 5.46 | 0.58 | 4.38 | 5.20 | 1.19 |
| Pd | 5.01 | 5.26 | 0.77 | 4.46 | 5.24 | 1.26 |
| Pt | 4.96 | 5.23 | 0.70 | 4.74 | 5.20 | 1.03 |

Table 4. DFT calculated effective work function (eWF) and electron Schottky barrier ($\Phi_B$) of contact metals with different strains δ (%) by D3 exchange functionals in Figure 4e. The supercell (SC) structure $\sqrt{m} \times \sqrt{m}$ Metal /$\sqrt{n} \times \sqrt{n}$ TMD is represented by m/n for simplicity.

| | $MoS_2$ | | | | $WSe_2$ | | | |
|---|---|---|---|---|---|---|---|---|
| | SC | δ | eWF | $\Phi_B$ | SC | δ | eWF | $\Phi_B$ |
| Bi | 9/19 | 1.1 | 4.34 | 0.12 | 7/13 | -1.1 | 3.88 | 0.38 |
| | 4/9 | 4.2 | 4.32 | 0.12 | 4/7 | -4.1 | 3.95 | 0.38 |
| Sb | 4/7 | -2.9 | 4.30 | 0.09 | 4/7 | 1.3 | 4.13 | 0.61 |
| | 7/13 | 0.1 | 4.28 | 0.08 | 7/12 | 0.3 | 4.05 | 0.57 |
| Au | 9/7 | -3.3 | 4.88 | 0.52 | 4/3 | -0.9 | 4.33 | 0.83 |
| | 16/13 | -1.1 | 4.63 | 0.44 | 21/16 | -0.1 | 4.20 | 0.77 |
| Pd | 4/3 | -0.4 | 5.01 | 0.80 | 4/3 | 3.7 | 4.47 | 0.98 |
| | 9/7 | 1.4 | 5.00 | 0.81 | 13/9 | -0.2 | 4.50 | 1.01 |
| Pt | 4/3 | -1.2 | 4.92 | 0.69 | 4/3 | 2.9 | 4.37 | 0.88 |
| | 9/7 | 0.6 | 4.98 | 0.80 | 13/9 | -1.0 | 4.51 | 1.03 |

## ■ METHODS

**X-ray and Ultraviolet Photoelectron Spectroscopy**:

X-ray photoelectron spectroscopy (XPS) and ultraviolet photoelectron spectroscopy (UPS) measurements were performed using a PHI spectrometer. For XPS, a monochromatic Al Kα source (hν = 1486.6 eV) was employed, providing an overall energy resolution better than 0.5 eV. Spectra were acquired at a photoelectron take-off angle of 45° under ultrahigh vacuum conditions (base pressure $< 1 \times 10^{-9}$ mbar). UPS measurements were carried out using a He I discharge lamp (hν = 21.22 eV) as the excitation source. The secondary electron cutoff was measured under a –5 V sample bias to accurately determine the work function. For both XPS and UPS, binding energies were referenced to the Fermi edge of a clean Au foil. All spectral data were processed using CasaXPS software, including background subtraction, peak fitting, and deconvolution, to extract quantitative chemical and electronic structure information.

High-resolution XPS spectra of W 4*f* and Mo 3*d* core levels for $WSe_2$ and $MoS_2$ with metal contacts (Pd, Pt, Bi, Au) were measured. Spectra were fitted using a Shirley background and Gaussian–Lorentzian line shapes (70% Gaussian) in CasaXPS.

**Scanning tunneling spectroscopy (STS):**

Scanning Tunneling Spectroscopy (STS) measurements were conducted at a constant current of 300 pA and a temperature of 300 K. The base pressure of the ultra-high vacuum (UHV) scanning tunneling microscope (STM) chamber was maintained below $3.0\times10^{-10}$ torr. An electrochemically etched tungsten tip, used for the measurements, was characterized and verified through test measurements performed on an Au (111) surface. Further calibration of the STM system was achieved using a bulk Highly Oriented Pyrolytic Graphite (HOPG) sample. The band gap energy was determined by linearly fitting the intersection points at the band gap edges, a method consistent with previous studies on metal/1L TMD systems.[35]

**First-principles calculations:**

**Bandstructure, wokfunction, and binding energy calculations:**

The first-principles calculations for the metal/$MoS_2$ and metal/$WSe_2$ were performed by the Vienna Ab initio Simulation Package (VASP)[36,37] based on the projector augmented wave (PAW) method[38] that uses pseudopotential operators but keeps the full all-electron wave function. The exchange-correlation potential was described by the Perdew–Burke–Ernzerhof generalized gradient approximation (PBE-GGA) approximation,[39] with the van der Waals correct ion incorporated by the vdW-DF (optB86)[40], DFT-D3,[41] vdW-DF2[42] functionals. The pseudopotential library was from the PseudoDojo library. The structure was fully relaxed until the change of energy and the force reached $10^{-6}$ eV/cell and $10^{-3}$ eV/Å, respectively. The optimized lattice constant for the $MoS_2$ and $WSe_2$ is 3.165 Å and 3.30 Å. All calculations were performed after relaxing the interlayer distance while keeping the lattice constants fixed. The geometry was set up by using the slab model and choosing the vacuum distance to be larger than 20 Å to eliminate the spurious interaction. The plane-wave energy cut-off is at least 450 eV. The k-points grids contain at least 144 points in a primitive Brillouin zone of $MoS_2$, $WS_2$, or $WSe_2$. The spin-orbit coupling (SOC) was not considered in the band structure calculations since the SOC effect is negligible in the conduction band for electron barrier height extraction. The metal film used for the simulations was restricted to four layers or a thickness of approximately 1 nm to manage computational costs. This constraint leads to varying work function (WF) values for the same metal on different transition metal dichalcogenides (TMDs) such as $MoS_2$, $WS_2$, and $WSe_2$. For instance, Sb contacts yielded WF values of 3.99, 4.00, and 4.35 eV on $MoS_2$, $WS_2$, and $WSe_2$, respectively, as illustrated in Figure 1b.

**Projected local density of states:**

The structure and local density of states (LDOS) of top contact Pd(111)/$MoS_2$ and edge contact Pd(111)-$MoS_2$ shown in Figures 7b and 7c were relaxed and calculated by Synopsys® atomistic simulation platform - "QuantumATK"[43] of PBE exchange-correlation functional with a Linear Combination of Atomic Orbitals (LCAO) basis set (FHI pseudopotentials with DoubleZetaPolarized (DZP) basis set). The k-point sampling is 18×1×95 k-point sampling with a periodic, Neumann, and Dirichlet boundary condition, respectively. Density mess cut-off is 60 Hartree energy.

## ■ ASSOCIATED CONTENT

**Supporting Information**

The Supporting Information is available free of charge at https://pubs.acs.org/doi/10.1021/acsnano.[Insert DOI].

Calculated Schottky barrier heights (SBHs) versus effective work function (eWF) for monolayer and bilayer $MoS_2$,, $WS_2$,, and $WSe_2$ using different exchange functionals (Figures S1 and S7); structural parameters, eWFs, and SBHs for metal-monolayer and metal-bilayer transition metal dichalcogenide (TMD) superlattices (Table S1 and S2); projected band structures illustrating orbital hybridization and metal-induced gap states for Hf/$MoS_2$ and Ti/$MoS_2$ interfaces (Figure S2); calculated binding energy profiles fitted with a modified 6-3 Lennard-Jones potential for various metal/TMD contacts along with an analysis of deviations in the chemisorption regime (Figure S3); X-ray photoelectron spectroscopy (XPS) core-level binding energy shifts for pristine and metal-deposited $WSe_2$ and $MoS_2$ samples (Figure S4); Moiré structures of Au/$WSe_2$ and Pd/$WSe_2$ containing single and multiple Se defects (Figure S5); and calculated eWF trends as a function of Se defect concentration for Pt, Pd, and Au contacts on $WSe_2$ (Figure S6)

## ■ REFERENCES

**1** Kroemer, H. "Nobel Lecture: Quasielectric fields and band offsets: teaching electrons new tricks". *Rev. Mod. Phys.* **2001**, *73*, 783, DOI: 10.1103/RevModPhys.73.783

**2** Novoselov, K. S.; Geim, A. K.; Morozov, S. V.; Jiang, D.; Zhang, Y.; Dubonos, S. V.; Grigorieva, I. V.; Firsov, A. A. "Electric field effect in atomically thin carbon films". *Science* **2004**, *306*, 666-669, DOI: 10.1126/science.1102896

**3** Bardeen, J. "Surface States and Rectification at a Metal Semi-Conductor Contact". *Phys. Rev.* **1947**, *71*, 717, DOI: 10.1103/PhysRev.71.717

**4** Heine, V. "Theory of Surface States". *Phys. Rev.* **1965**, *138*, A1689, DOI: 10.1103/PhysRev.138.A1689

**5** Tung, R. T. "Formation of an electric dipole at metal-semiconductor interfaces". *Phys. Rev. B* **2001**, *64*, 205310, DOI: 10.1103/PhysRevB.64.205310

**6** Gong, C.; Colombo, L.; Wallace, R. M.; Cho, K. "The Unusual Mechanism of Partial Fermi Level Pinning at Metal–MoS2 Interfaces". *Nano Lett.* **2014**, *14*(4), 1714-1720, DOI: 10.1021/nl403465v

**7** Farmanbar, M.; Brocks, G. "Controlling the Schottky barrier at MoS2/metal contacts by inserting a BN monolayer". *Phys. Rev. B* **2015**, *91*, 161304, DOI: 10.1103/PhysRevB.91.161304

**8** Farmanbar, M.; Brocks, G. "First-principles study of van der Waals interactions and lattice mismatch at MoS2/metal interfaces". *Phys. Rev. B* **2016**, *93*, 085304, DOI: 10.1103/PhysRevB.93.085304

**9** Gong, M.; Xie, D.; Tian, Y.; Hua, Z.; Zhang, C.; Li, M.; Cao, D.; Zhou, J.; Chen, X.; Shu, H. "Unraveling the Role of Interfacial Interactions in Electrical Contacts of Atomically Thin Transition-Metal Dichalcogenides". *ACS Nano* **2025**, *19*(4), 4718-4730, DOI: 10.1021/acsnano.4c15341

**10** Deng, H.-X.; Zhang, C.; Xiao, J.; Yang, K.-K.; Qiu, C.; Wei, Z.; Liu, Y.-Y.; Luo, J.-W.; Wei, S.-H. "Compatibility between Schottky–Mott Limit and High Carrier-Injection Efficiency in Metal–van der Waals Semiconductor Junctions". *Nano Lett.* **2025**, *25*(41), 14880-14886, DOI: 10.1021/acs.nanolett.5c03335

**11** Das, S.; Chen, H.-Y.; Penumatcha, A. V.; Appenzeller, J. "High Performance Multilayer MoS2 Transistors with Scandium Contacts". *Nano Lett.* **2013**, *13*(1), 100-105, DOI: 10.1021/nl303583v

**12** Kim, C.; Moon, I.; Lee, D.; Choi, M. S.; Ahmed, F.; Nam, S.; Cho, Y.; Shin, H.-J.; Park, S.; Yoo, W. J. "Fermi Level Pinning at Electrical Metal Contacts of Monolayer Molybdenum Dichalcogenides". *ACS Nano* **2017**, *11*(2), 1588-1596, DOI: DOI:

10.1021/acsnano.6b07159

**13** Bampoulis, P.; van Bremen, R.; Yao, Q.; Poelsema, B.; Zandvliet, H. J. W.; Sotthewes, K. "Defect Dominated Charge Transport and Fermi Level Pinning in MoS2/Metal Contacts". *ACS Appl. Mater. Interfaces* **2017**, *9*(22), 19278-19286, DOI: 10.1021/acsami.7b02739

**14** Schulman, D. S.; Arnold, A. J.; Das, S. "Contact engineering for 2D materials and devices". *Chem. Soc. Rev.* **2018**, *47*, 3037-3058, DOI: 10.1039/C7CS00828G

**15** Sotthewes, K.; van Bremen, R.; Dollekamp, E.; Boulogne, T.; Nowakowski, K.; Kas, D.; Zandvliet, H. J. W.; Bampoulis, P. "Universal Fermi-Level Pinning in Transition-Metal Dichalcogenides". *J. Phys. Chem. C* **2019**, *123*(9), 5411-5420, DOI: 10.1021/acs.jpcc.8b10971

**16** Bampoulis, P.; van Bremen, R.; Yao, Q.; Poelsema, B.; Zandvliet, H. J. W.; Sotthewes, K. "Defect Dominated Charge Transport and Fermi Level Pinning in MoS2/Metal Contacts". *ACS Appl. Mater. Interfaces* **2017**, *9*(22), 19278-19286, DOI: 10.1021/acsami.7b02739

**17** Mitterreiter, E.; Schuler, B.; Micevic, A.; Hernangómez-Pérez, D.; Barthelmi, K.; Cochrane, K. A.; Kiemle, J.; Sigger, F.; Klein, J.; Wong, E.; Barnard, E. S.; Watanabe, K.; Taniguchi, T.; Lorke, M.; Jahnke, F.; Finley, J. J.; Schwartzberg, A. M.; Qiu, D. Y.; Refaely-Abramson, S.; Holleitner, A. W.; Weber-Bargioni, A.; Kastl, C. "The role of chalcogen vacancies for atomic defect emission in MoS2." *Nat. Commun.* **2021**, *12*, 3822, DOI: 10.1038/s41467-021-24102-y

**18** Noh, J.-Y.; Kim, H.; Kim, Y.-S. "Stability and electronic structures of native defects in single-layer MoS2". *Phys. Rev. B* **2014**, *89*, 205417, DOI: 10.1103/PhysRevB.89.205417

**19** Komsa, H.-P.; Krasheninnikov, A. V. "Native defects in bulk and monolayer MoS2

from first principles". *Phys. Rev. B* **2015**, *91*, 125304, DOI: 10.1103/PhysRevB.91.125304

**20** Cao, R.; Guo, H.; Luo, J.-W.; Guo, Y.; Robertson, J. "Density Functional Analysis of Threshold Voltage Control by Oxide Dipole Layers in Si- and MoS2-Based FETs". *ACS Appl. Electron. Mater.* **2025**, *7*(2), 721-728, DOI: 10.1021/acsaelm.4c01790

**21** Liu, Y.; Guo, J.; Zhu, E.; Liao, L.; Lee, S.-J.; Ding, M.; Shakir, I.; Gambin, V.; Huang, Y.; Duan, X. "Approaching the Schottky–Mott limit in van der Waals metal–semiconductor junctions". *Nature* **2018**, *557*, 696-700, DOI: 10.1038/s41586-018-0129-8

**22** Wang, Q.; Shao, Y.; Gong, P.; Shi, X. "Metal–2D multilayered semiconductor junctions: layer-number dependent Fermi-level pinning". J. Mater. Chem. C 2020, 8, 3113-3119, DOI: 10.1039/C9TC06331E

**23** Das, S.; Chen, H.-Y.; Penumatcha, A. V.; Appenzeller, J. "High Performance Multilayer MoS2 Transistors with Scandium Contacts". *Nano Lett.* **2013**, *13*(1), 100-105, DOI: 10.1021/nl303583v

**24** Yang, N.; Lin, Y. C.; Chuu, C.-P.; Rahman, M. S.; Wu, T.; Chou, A.-S.; Chen, H.-Y.; Woon, W.-Y.; Liao, S. S.; Huang, S.; Qian, X.; Guo, J.; Radu, I.; Wong, H.-S. P.; Wang, H. "Ab initio computational screening and performance assessment of van der Waals and semimetallic contacts to monolayer WSe2 p-type field-effect transistors". *IEEE Trans. Electron Devices* **2023**, *70*(4), 2090-2097. DOI: 10.1109/TED.2023.3241569

**25** Shen, T.; Ren, J.-C.; Liu, X.; Li, S.; Liu, W. "van der Waals Stacking Induced Transition from Schottky to Ohmic Contacts: 2D Metals on Multilayer InSe". *J. Am. Chem. Soc.* **2019**, *141*(7), 3110-3115, DOI: 10.1021/jacs.8b12212

**26** Huang, A.; The, Y.-H.; Chen, C.-H.; Hung, S.-H.; Wang, J.-F.; Chuu, C.-P.; Jeng,

H.-T. "Tunable Work Function and Surface Energy in Titanium Nitride (TiN) Thin Films through Quantum Well States". *ACS Mater. Au* **2025**, *5*(2), 430-437, DOI: 10.1021/acsmaterialsau.4c00176

**27** Zuki, F. M.; Edyvean, R. G. J.; Pourzolfaghar, H.; Kasim, N. "Modeling of the Van Der Waals Forces during the Adhesion of Capsule-Shaped Bacteria to Flat Surfaces". *Biomimetics* **2021**, *6(1)*, 5, DOI: 10.3390/biomimetics6010005

**28** Rokni, H.; Lu, W. Direct measurements of interfacial adhesion in 2D materials and van der Waals heterostructures in ambient air. *Nat. Commun.* **2020**, *11*, 5607, DOI: 10.1038/s41467-020-19411-7

**29** Smyth, C. M.; Walsh, L. A.; Bolshakov, P.; Catalano, M.; Addou, R.; Wang, L.; Kim, J.; Kim, M. J.; Young, C. D.; Hinkle, C. L.; Wallace, R. M. Engineering the interface chemistry for scandium electron contacts in WSe2 transistors and diodes. *2D Mater.* **2019**, *6*, 045020, DOI: 10.1088/2053-1583/ab2c44

**30** Smyth, C. M.; Walsh, L. A.; Bolshakov, P.; Catalano, M.; Addou, R.; Wang, L.; Kim, J.; Kim, M. J.; Young, C. D.; Hinkle, C. L.; Wallace, R. M. Engineering the palladium–WSe2 interface chemistry for field effect transistors with high-performance hole contacts. *ACS Appl. Nano Mater.* **2019**, *2*, 75–88, DOI: 10.1021/acsanm.8b01708

**31** Kwon, G.; Choi, Y.-H.; Lee, H.; Kim, H.-S.; Jeong, J.; Jeong, K.; Baik, M.; Kwon, H.; Ahn, J.; Lee, E.; Cho, M.-H. Interaction- and defect-free van der Waals contacts between metals and two-dimensional semiconductors. *Nat. Electron.* **2022**, *5*, 241–247, DOI: 10.1038/s41928-022-00746-6

**32** Addou, R.; McDonnell, S.; Barrera, D.; Guo, Z.; Azcatl, A.; Wang, J.; Zhu, H.; Hinkle, C. L.; Quevedo-Lopez, M.; Alshareef, H. N.; Colombo, L.; Hsu, J. W. P.; Wallace, R. M. Impurities and Electronic Property Variations of Natural MoS2 Crystal Surfaces. *ACS Nano* **2015**, *9*, 9124–9133, DOI: 10.1021/acsnano.5b03309

**33** Miller, T.; Chou, M. Y.; Chiang, T.-C. Phase relations associated with one-dimensional shell effects in thin metal films. *Phys. Rev. Lett.* **2009**, *102*, 236803, DOI: https://doi.org/10.1103/PhysRevLett.102.236803

**34** Hung, Terry Y. T.; Wang, S.-Y.; Chuu, C.-P.; Chung, Y.-Y.; Chou, A.-S.; Huang, F.-S.; Chen, T.; Li, M.-Y.; Cheng, C.-C.; Cai, J.; Chien, C.-H.; Chang, W.-H.; Wong, H.-S. P.; Li, L.-J. Pinning-free edge contact monolayer MoS2 FET. *IEEE International Electron Devices Meeting (IEDM)*; Institute of Electrical and Electronics Engineers, **2020**; 3.3.1–3.3.4. DOI: 10.1109/IEDM13553.2020.9372028

**35** Chen, Y.-F.; Hsu, H.-C.; Chen, H.-Y.; Chen, L.-Y.; Lin, Y.-R.; Li, M.-Y.; Radu, I. P.; Chiu, Y.-P. Direct visualization of metal-induced gap state distribution and valley band evolution at metal versus semimetal MoS2 interfaces. *ACS Nano* **2025**, *19*, 19408–19416, DOI: 10.1021/acsnano.5c03676

**36** Kresse, G.; Furthmuller, J. Efficiency of ab-initio total energy calculations for metals and semiconductors using a plane-wave basis set. *Comput. Mater. Sci.* **1996**, *6*, 15, DOI: 10.1016/0927-0256(96)00008-0

**37** Kresse, G.; Furthmuller, J. Efficient iterative schemes for ab initio total-energy calculations using a plane-wave basis set. *Phys. Rev. B* **1996**, *54*, 11169, DOI: 10.1103/PhysRevB.54.11169

**38** Blöchl, P. E. Projector augmented-wave method. *Phys. Rev. B* **1994**, *50*, 17953, DOI: 10.1103/PhysRevB.50.17953

**39** Perdew, J. P.; Burke, K.; Ernzerhof, M. Generalized Gradient Approximation made simple. *Phys. Rev. Lett.* **1996**, *77*, 3865–3868, DOI: 10.1103/PhysRevLett.77.3865

**40** Klimeš, J.; Bowler, D. R.; Michaelides, A. Van der Waals density functionals applied to solids. *Phys. Rev. B* **2011**, *83*, 195131, DOI: 10.1103/PhysRevB.83.195131

**41** Grimme, S.; Antony, J.; Ehrlich, S.; Krieg, H. A consistent and accurate ab initio

parametrization of density functional dispersion correction (DFT-D) for the 94 elements H-Pu. *J. Chem. Phys.* **2010**, *132*, 154104, DOI: 10.1063/1.3382344

**42** Lee, K.; Murray, É. D.; Kong, L.; Lundqvist, B. I.; Langreth, D. C. Higher-accuracy van der Waals density functional. *Phys. Rev. B* **2010**, *82*, 081101, DOI: 10.1103/PhysRevB.82.081101

**43** Smidstrup, S.; Markussen, T.; Vancraeyveld, P.; Wellendorff, J.; Schneider, J.; Gunst, T.; Verstichel, B.; Stradi, D.; Khomyakov, P. A.; Vej-Hansen, U. G.; Lee, M.-E.; Chill, S. T.; Rasmussen, F.; Penazzi, G.; Corsetti, F.; Ojanperä, A.; Jensen, K.; Palsgaard, M. L. N.; Martinez, U.; Blom, A.; Brandbyge, M.; Stokbro, K. Quantumatk: An integrated platform of electronic and atomic-scale modelling tools. *J. Phys.: Condens. Matter* **2020**, *32*, 015901, DOI: 10.1088/1361-648X/ab4007

# Supporting Information

## Revisiting the Schottky-Mott Relation for Metal/Transition Metal Dichalcogenide Interfaces: The Role of Interface Dipoles

Chih-Piao Chuu[1*], Shu-Jui Chang[1], Yu-Miin Sheu[3], Zih-Syuan Huang[1,2], Wei-Tung Liu [3], Po-Wei Chen[3], Sui-An Chou[1], Yu-Tung Lin[1,2], Chih-I Wu[2], Wen-Hao Chang[3], Chun-Liang Lin[3], H.-S. Philip Wong[4], and Iuliana P. Radu[1*]

[1] Corporate Research, Taiwan Semiconductor Manufacturing Company, Hsinchu 30091, Taiwan;

[2] Graduate Institute of Photonics and Optoelectronics, National Taiwan University, Taipei 106, Taiwan;

[3] Department of Electrophysics, National Yang Ming Chiao Tung University, 1001 Ta Hsueh Road, Hsinchu 30050, Taiwan;

[4] Department of Electrical Engineering and Stanford SystemX Alliance Paul G. Allen Building 312X 420 Via Palou, Stanford University, Stanford, California 94305, United State;

* Corresponding author: *Email: cpchuu@tsmc.com (C.-P. Chuu); Email: IRADU@tsmc.com (I. P. Radu)*

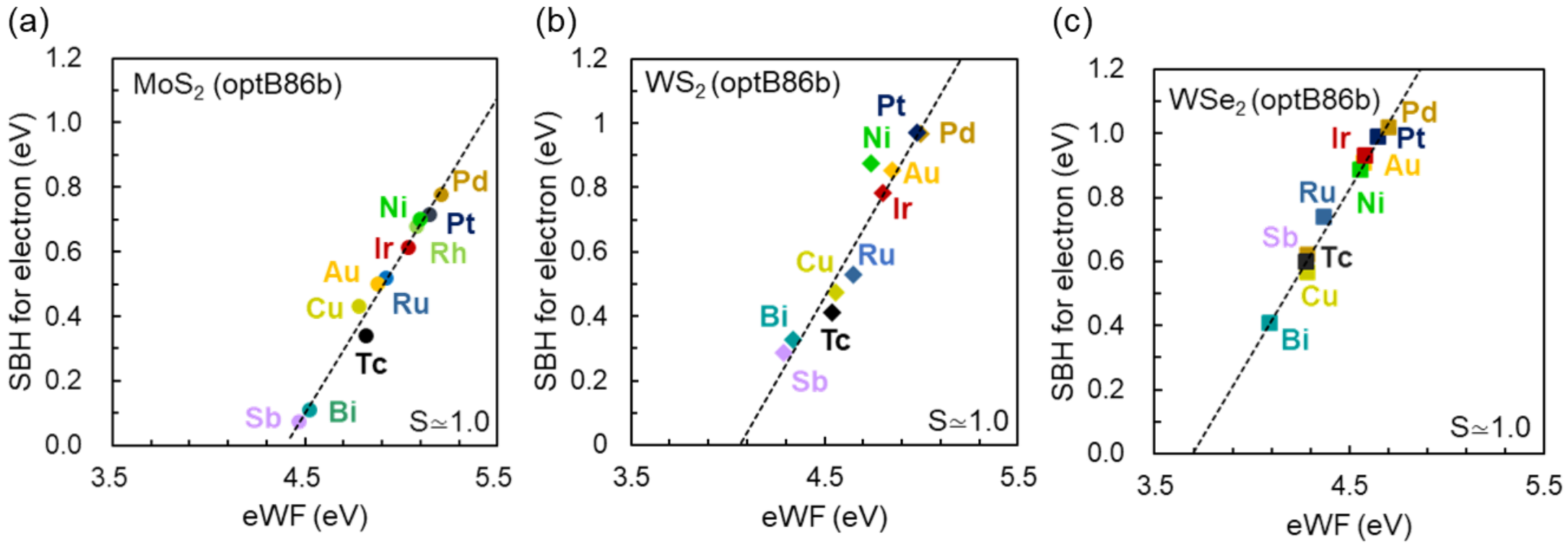

Figure S1. Calculated Schottky barrier heights (SBHs) of electrons v.s. effective work function (eWF) of contact metals on monolayer (a) $MoS_2$ (circles), (b) $WS_2$ (diamonds), and (c) $WSe_2$ (squares) using different exchange functionals optB86b. The details of calculations are listed in Table S1.

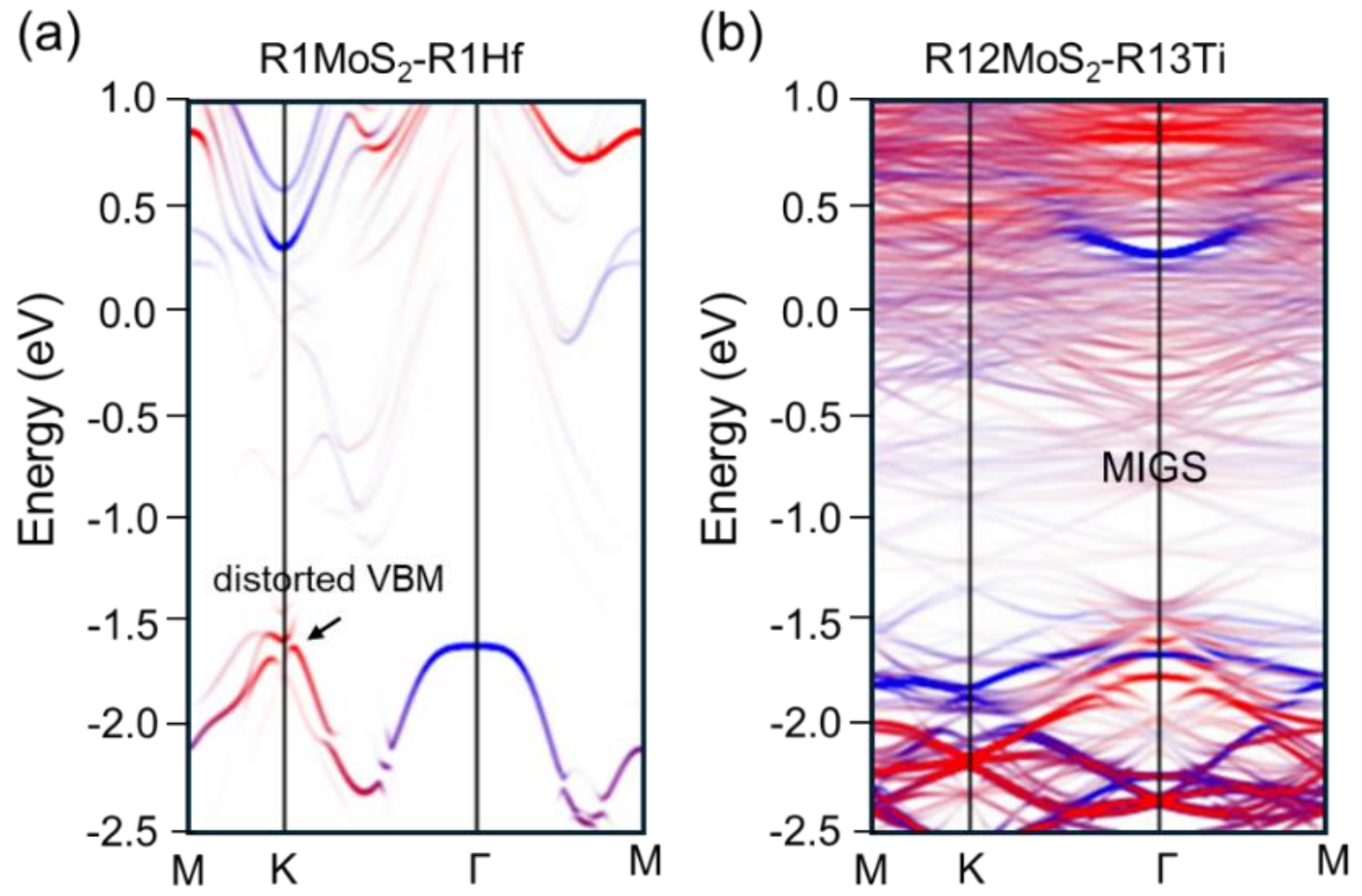

Figure S2. Calculated projected band structures for R1$MoS_2$-R1Hf and R12$MoS_2$-R13Ti. The $d_{z^2}$ orbitals of Mo atoms are highlighted in red, while $d_{x^2-y^2}/d_{xy}$ orbitals of Mo atoms are shown in blue. In the case of Hf/$MoS_2$, strong interlayer interaction leads to a distortion of the VBM at the K valley. For Ti/$MoS_2$, a dense MIGS sea is observed in the midgap region.

Modified 6-3 Lennard-Jones potential:

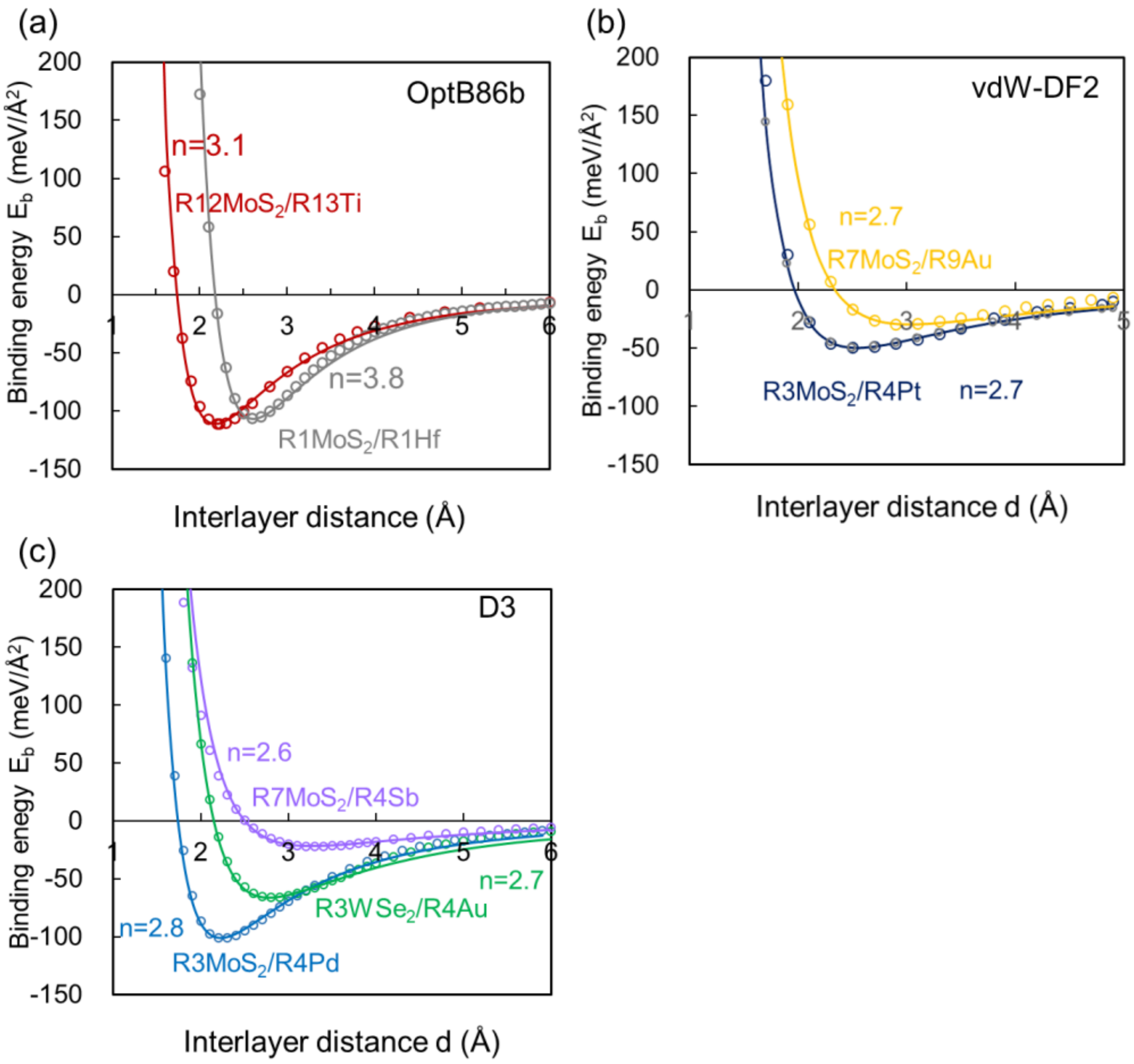

Figure S3. (a) Calculated binding energy ($E_b$, circles) for Ti/$MoS_2$ (R13/R12, magenta) and Hf/$MoS_2$ (R1/R1, grey) with optB86b exchange functionals. These data are fitted with a 6-3 Lennard-Jones potential (solid line) using powers n=3.1 and n=3.8 for Ti/$MoS_2$ and Hf/$MoS_2$, respectively. (b) Calculated binding energy ($E_b$, circles) for Au/$MoS_2$ (R7/R9, orange) and Pt/$MoS_2$ (R3/R4, navy) with vdW-DF2 exchange functionals, fitted with a 6-3 L-J potential (solid line) of powers n=2.7 for both. (c) Calculated binding energy ($E_b$, circles) for Sb/$MoS_2$ (R7/R4, purple), Au/$WSe_2$ (R3/R4, green), Pd/$MoS_2$ (R3/R4, blue) with vdW-DF2 exchange functionals, fitted with a 6-3 L-J potential (solid line) of powers n=2.6, 2.7, and 2.8 for Sb/$MoS_2$, Au/$WSe_2$, and Pd/MoS2, respectively.

The Lennard-Jones (LJ) potential, a standard empirical model for van der Waals (vdW) interactions, inherently assumes non-covalent forces characteristic of physisorption. This is consistent with our DFT binding-energy calculations, which reveal significant deviations from the fitted LJ potentials when the interlayer distance (d_int) falls below $d_{opt}$ -0.5 Å. This repulsive regime is typically associated with strong orbital overlap and charge transfer, which critically influence the effective work function (eWF) and Schottky barrier height (SBH) as a function of interlayer distance (see Figures 6a–c).

While the LJ potential accurately captures the attractive region ($d_{int} > d_{opt}$ - 0.5 Å), consistent with the Hamaker approach, its limitations become evident in the strong chemisorption regime. For example, Ti and Hf exhibit large binding energies at equilibrium (Table 2), together with pronounced band distortion (Hf/$MoS_2$) and prominent metal-induced gap states (Ti/$MoS_2$) in their projected band structures (Figure S2). These features clearly indicate strong orbital hybridization and charge transfer, characteristic of chemisorptive rather than vdW bonding.

The inadequacy of a purely vdW description is further evidenced by deviations between the fitted LJ potentials and DFT results, reaching ~3–4 meV/Å$^2$ (Ti/$MoS_2$) and 2–3 meV/Å$^2$ (Hf/$MoS_2$), even at relatively large separations. Consequently, such chemisorptive interlayer interactions pose substantial challenges for both theoretical approaches (e.g., effective work function models) and experimental techniques (e.g., scanning tunneling spectroscopy) in determining band-edge positions, as band tails may obscure the precise identification of band edges and Schottky barrier heights.

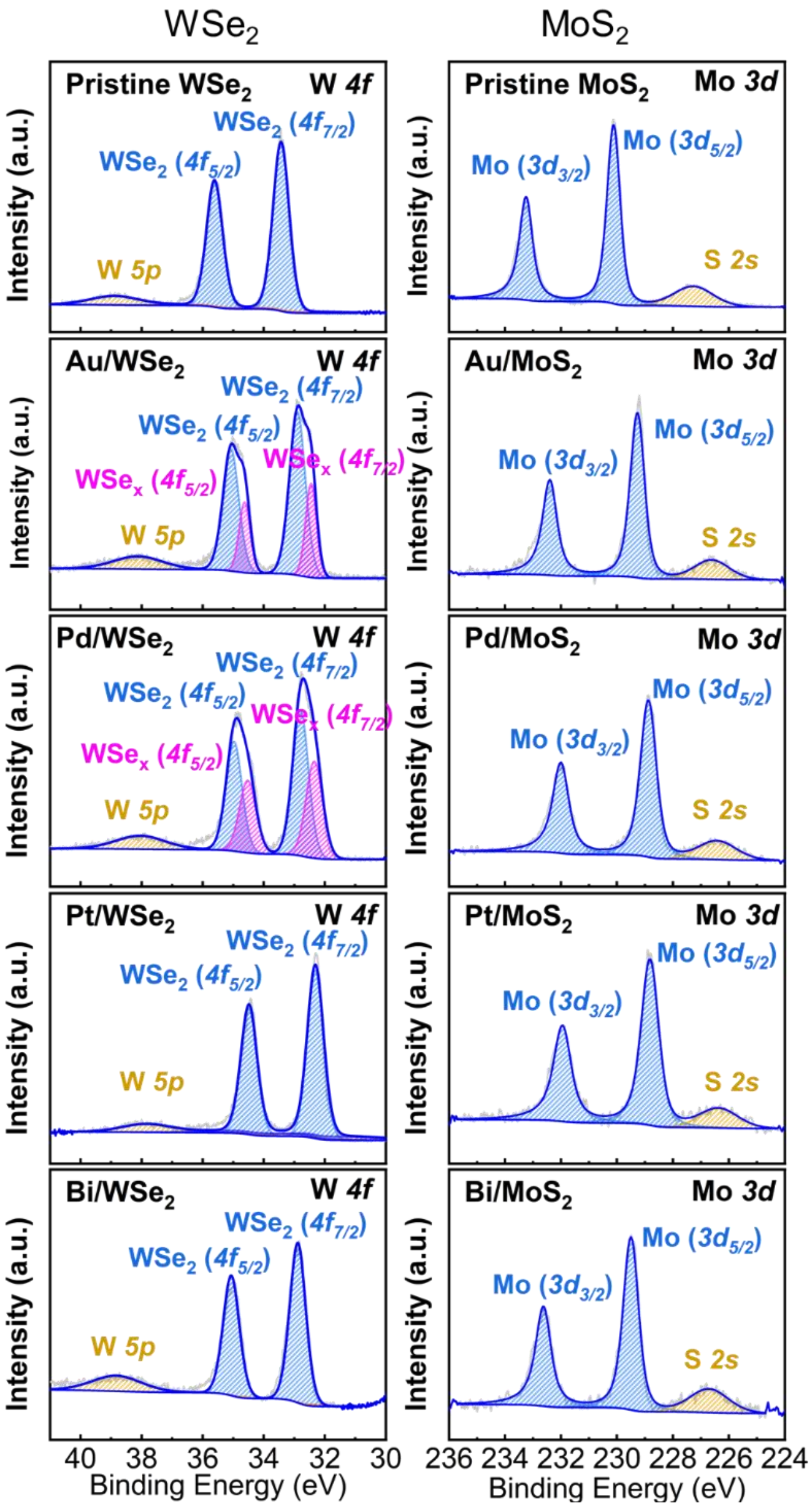

Figure S4. The XPS binding energy measurement of $WSe_2$ and $MoS_2$ samples before metal contact (pristine) and after metal deposited metal/TMD. The core levels have shifted in both $WSe_2$ and $MoS_2$ samples after metal deposits. Doublets of peak were only observed in Au/$WSe_2$ and Pt/$WSe_2$ samples.

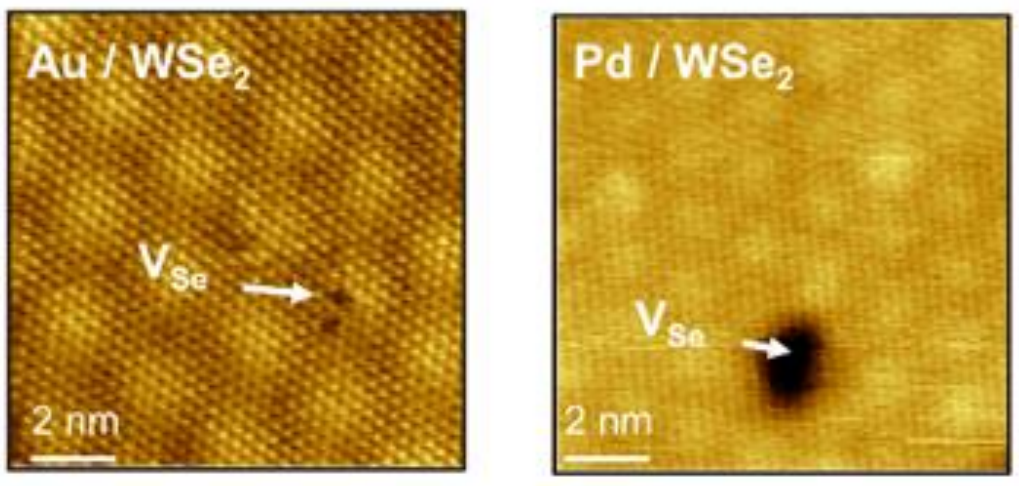

Figure S5. Moiré structures of Au/$WSe_2$ and Pd/$WSe_2$ with single and multiple Se defects

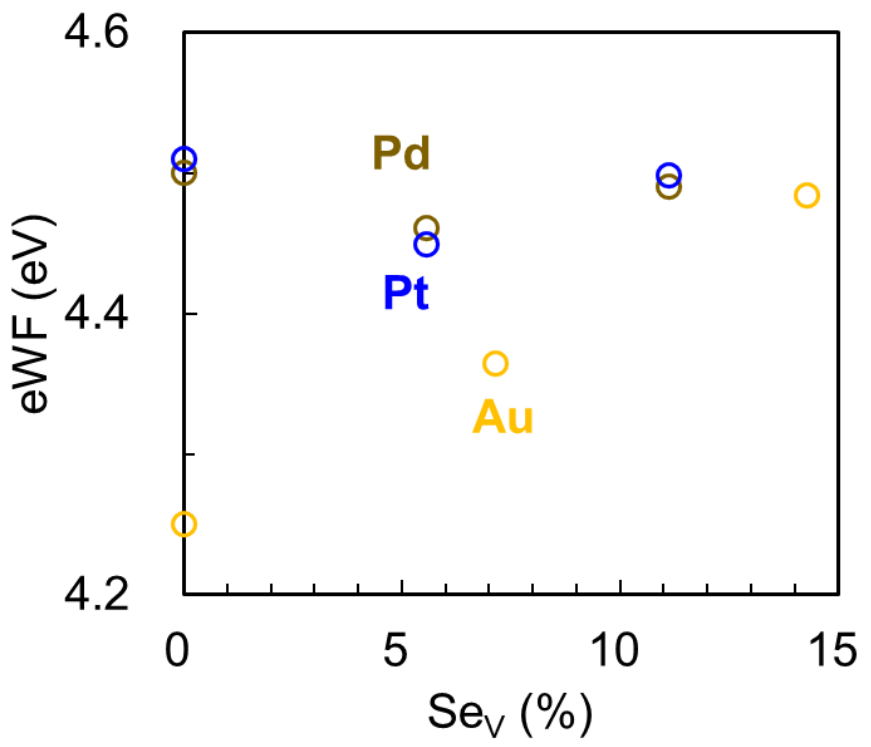

Figure S6. Calculated eWF versus Se defect concentration (%) for $Pt/WSe_2$, $Pd/WSe_2$, and $Au/WSe_2$.

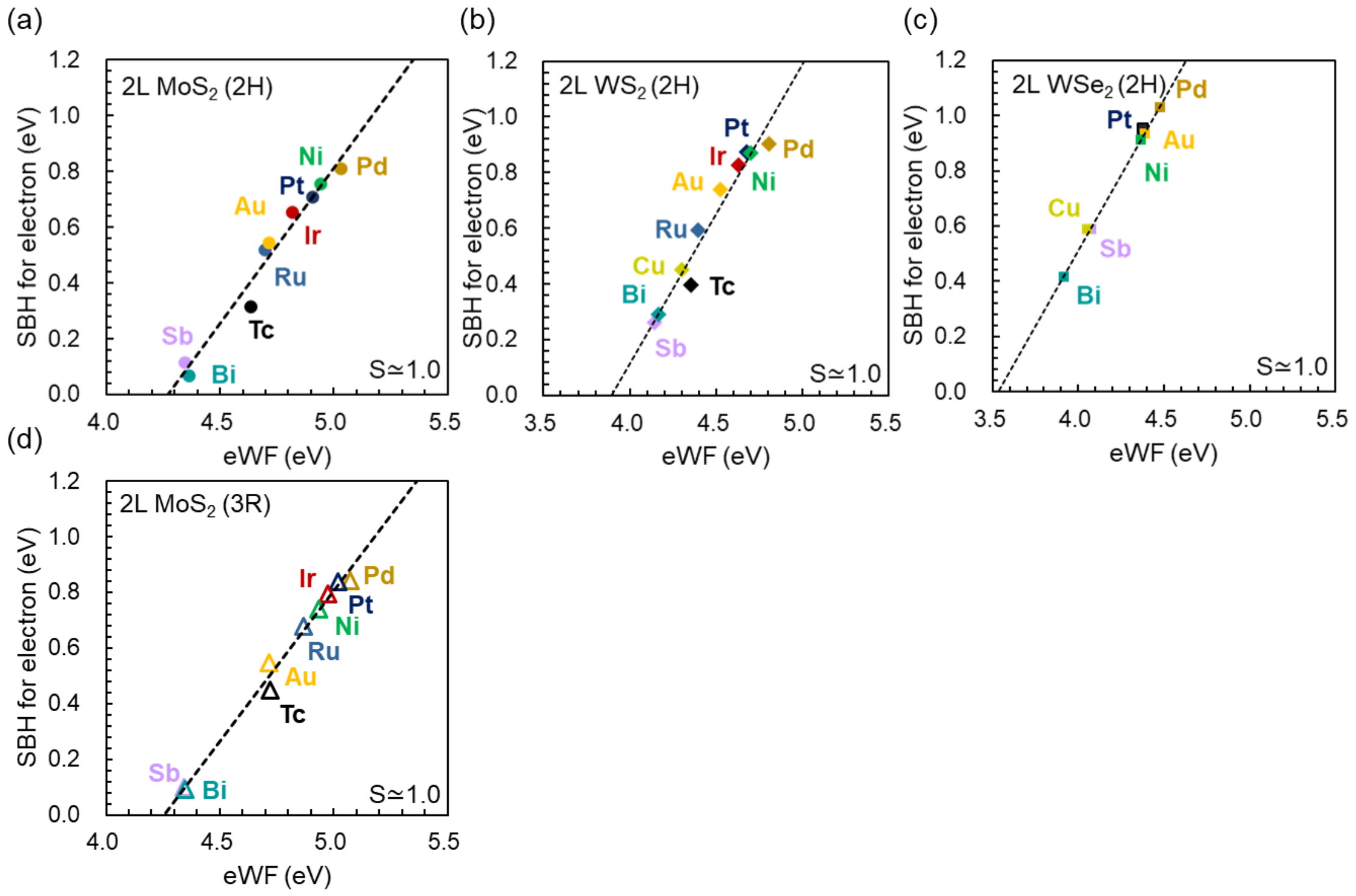

Figure S7. Calculated Schottky barrier heights (SBHs) of electrons v.s. effective work function (eWF) of contact metals on bilayer $MoS_2$ (circles), bilayer $WS_2$ (diamonds), and bilayer $WSe_2$ (squares). Slope S was obtained from the linear fitting (solid line) of calculated data points of SBH versus eWF for bilayer $MoS_2$, bilayer $WS_2$, and bilayer $WSe_2$. Calculation details are listed in Table S2 below.

Table S1. Calculated interlayer distance $d_{int}$, effective work function (eWF) and electron Schottky barrier $\Phi_B$ of metal-monolayer TMD superlattices structure by optB86b exchange functionals in Figure S1. The supercell (SC) structure $\sqrt{m} \times \sqrt{m}$ Metal /$\sqrt{n} \times \sqrt{n}$ TMD is represented by m/n for simplicity. The calculated lattice constant of monolayer $MoS_2$, $WS_2$, and $WSe_2$ is 3.169 Å, 3.169 Å, 3.30 Å, respectively.

| | $MoS_2$ | | | | $WS_2$ | | | | $WSe_2$ | | | |
|---|---|---|---|---|---|---|---|---|---|---|---|---|
| optB86b | SC | $d_{int}$ | eWF | $\Phi_B$ | SC | $d_{int}$ | eWF | $\Phi_B$ | SC | $d_{int}$ | eWF | $\Phi_B$ |
| Pt | 4/3 | 2.50 | 5.15 | 0.72 | 4/3 | 2.68 | 4.98 | 0.97 | 4/3 | 2.58 | 4.65 | 0.99 |
| Pd | 4/3 | 2.27 | 5.21 | 0.78 | 4/3 | 2.36 | 5.00 | 0.96 | 4/3 | 2.42 | 4.70 | 1.02 |
| Bi | 9/19 | 3.13 | 4.52 | 0.11 | 9/19 | 3.24 | 4.34 | 0.33 | 12/13 | 3.35 | 4.09 | 0.41 |
| Sb | 4/7 | 3.30 | 4.47 | 0.07 | 4/7 | 3.45 | 4.30 | 0.28 | 7/4 | 3.40 | 4.28 | 0.60 |
| Ni | 7/4 | 2.10 | 5.10 | 0.70 | 7/4 | 2.14 | 4.85 | 0.85 | 7/4 | 2.28 | 4.56 | 0.88 |
| Au | 9/7 | 2.78 | 4.78 | 0.50 | 9/7 | 3.05 | 4.74 | 0.87 | 4/3 | 3.00 | 4.58 | 0.91 |
| Ir | 4/3 | 2.40 | 5.04 | 0.61 | 4/3 | 2.42 | 4.81 | 0.78 | 13/9 | 2.60 | 4.58 | 0.93 |
| Cu | 13/9 | 2.30 | 4.78 | 0.43 | 19/12 | 2.60 | 4.56 | 0.47 | 7/4 | 2.59 | 4.29 | 0.56 |
| Ru | 4/3 | 2.27 | 4.92 | 0.52 | 4/3 | 2.27 | 4.65 | 0.53 | 13/9 | 2.43 | 4.37 | 0.74 |
| Tc | 4/3 | 2.34 | 4.82 | 0.34 | 4/3 | 2.38 | 4.54 | 0.41 | 13/9 | 2.53 | 4.28 | 0.62 |
| Rh | 4/3 | 2.20 | 5.08 | 0.68 | | | | | | | | |

Table S2. Calculated effective work function (eWF) and electron Schottky barrier $\Phi_B$ of metal-bilayer $MoS_2$, $WS_2$, and $WSe_2$ supercell by D3 exchange functionals. Identical supercells and interlayer distance of $\sqrt{m} \times \sqrt{m}$ Metal /$\sqrt{n} \times \sqrt{n}$ TMD were adopted for bilayer calculations.

| | 2L $MoS_2$ | | | | 2L $WS_2$ | | 2L $WSe_2$ | |
|---|---|---|---|---|---|---|---|---|
| | 2H | | 3R | | 2H | | 2H | |
| | eWF | $\Phi_B$ | eWF | $\Phi_B$ | eWF | $\Phi_B$ | eWF | $\Phi_B$ |
| Pd | 5.03 | 0.81 | 5.08 | 0.84 | 4.81 | 0.90 | 4.48 | 1.03 |
| Pt | 4.91 | 0.71 | 5.02 | 0.84 | 4.68 | 0.87 | 4.38 | 0.95 |
| Au | 4.72 | 0.54 | 4.72 | 0.55 | 4.53 | 0.74 | 4.39 | 0.93 |
| Ni | 4.94 | 0.76 | 4.94 | 0.74 | 4.70 | 0.87 | 4.37 | 0.91 |
| Bi | 4.37 | 0.07 | 4.35 | 0.09 | 4.17 | 0.29 | 3.92 | 0.41 |
| Sb | 4.35 | 0.11 | 4.34 | 0.10 | 4.15 | 0.26 | 4.08 | 0.59 |
| Ir | 4.82 | 0.65 | 4.97 | 0.80 | 4.64 | 0.82 | | |
| Ru | 4.70 | 0.52 | 4.87 | 0.68 | 4.40 | 0.59 | | |
| Tc | 4.64 | 0.32 | 4.72 | 0.45 | 4.36 | 0.39 | | |
| Cu | | | | | 4.31 | 0.45 | 4.06 | 0.59 |